\documentclass [preprint,aps,showpacs]{revtex4}
\def\be{\begin{eqnarray} &&}
\def\nonu{\nonumber \\ &&}
\def\ee{\end{eqnarray}}
\def\psla{ \slash \! \! \!}
\def\Psla{ \slash \!\! \! \!}

\usepackage{graphicx}
\usepackage{epsfig}
\usepackage{amssymb}
\begin{document}

\title{Light-front Ward-Takahashi Identity for Two-Fermion Systems}

\author{J. A. O. Marinho $^a$, T. Frederico$^a$, E. Pace$^b$,
G. Salm\`e$^c$ and P. U. Sauer$^d$} \affiliation{ $^a$ Dep. de F\'\i
sica, Instituto Tecnol\'ogico de Aeron\'autica,  12.228-900 S\~ao
Jos\'e dos
Campos, S\~ao Paulo, Brazil\\
$^b$ Dipartimento di Fisica, Universit\`a di Roma "Tor Vergata"
and Istituto Nazionale di Fisica Nucleare, Sezione Tor Vergata,
Via della Ricerca
Scientifica 1, I-00133  Roma, Italy \\
$^c$Istituto  Nazionale di Fisica Nucleare, Sezione Roma I, P.le A.
Moro 2, I-00185 Roma, Italy \\
$^d$Institute for Theoretical Physics, Leibniz University, D-30167
Hannover, Germany }

\begin{abstract}
We propose a three-dimensional electromagnetic current operator
within light-front dynamics that satisfies a light-front
Ward-Takahashi identity for two-fermion
systems. The light-front current operator
is obtained by a quasi-potential reduction of the four-dimensional
current operator  and  acts on the
light-front valence component of bound or scattering states.
A relation between the light-front valence wave function and the
four-dimensional Bethe-Salpeter amplitude both for bound or
scattering states is also derived, such that the matrix elements of the
four-dimensional current
operator can be fully recovered from the corresponding light-front
ones.
The light-front current operator can be
perturbatively calculated through a quasi-potential expansion,
 and the divergence of the proposed
current satisfies a Ward-Takahashi identity at any given
order of the expansion.
In the quasi-potential expansion the instantaneous
terms of the fermion propagator are accounted for by the effective
interaction and two-body currents. We exemplify our theoretical
construction in the Yukawa model in the ladder approximation, investigating
in detail the
current operator at the lowest nontrivial order of the quasi-potential
expansion  of the Bethe-Salpeter equation. The explicit
realization of the light-front form of the Ward-Takahashi identity
is verified. We also show the relevance of instantaneous terms and of
the pair contribution to the two-body current and the Ward-Takahashi
identity.
\end{abstract}

 \maketitle


\section{Introduction}

%
A detailed investigation of the electromagnetic (em) properties  of
hadrons (including also nuclei) requests a careful treatment of the
current operator for interacting-fermion systems. In particular the
gauge symmetry plays the well-known essential role, that formally
leads to the fulfillment of the Ward-Takahashi identity (WTI). {
Following the seminal paper by Gross and Riska \cite{gross}}, in a
fully covariant theory, namely in a field theoretical framework in
its full glory, one should first solve the Bethe-Salpeter (BS)
equation and obtain a consistent current operator; then through a
Mandelstam~\cite{mandelstan} approach, one should evaluate the
hadron em properties. Unfortunately BS equation can be solved only
within approximate schemes or for simple examples (see e.g.
\cite{tjon,grossbook,carbonellepja,alkofer,itz}).   { For instance,
an appealing} technique for solving the BS equation is { based on}
the quasipotential (QP) expansion { of the four dimensional $T$
matrix } (see, e.g. \cite{wolja}), where one introduces an auxiliary
 four-dimensional  free Green's function { (in \cite{wolja}
it was three-dimensional), hopefully clever enough to allow a
meaningful truncation of the expansion.  Then, one projects the
relevant operators onto a three-dimensional hyperplane for obtaining
a three-dimensional integral equation for both scattering and bound
states. These three-dimensional states allow  to fully reconstruct
the four-dimensional BS amplitude solution after applying proper {
reverse} projection operators \cite{sales00,sales01}}. The method is
in principle exact, giving the solution of the four-dimensional BS
equation if the QP is evaluated without any approximation. { As
shown in \cite{hierareq} for massive particles, with a proper choice
of the auxiliary free Green's function the QP expansion can be put
in correspondence with the decomposition of the wave function in
terms of Fock components on the light-front (LF) hyperplane, i. e.
on the three-dimensional space adopted in this paper}. There are
extended reviews~\cite{kp,karmaprep,brodsky} illustrating the
relevant features of the LF framework, introduced in a famous paper
by Dirac \cite{dirac}, and the interested reader can profit of
those.

The present paper is a part of a program which attempts: i) to solve
the four-dimensional BS equation for two interacting particles, both
bosons and fermions, by using the QP expansion and a LF
three-dimensional projection onto the LF hypersurface
(see~\cite{sales00,sales01,adnei07} for  previous works) and, ii) to
develop a procedure for evaluating matrix elements of the
four-dimensional em current keeping WTI valid, at any order in the
truncation. { A key role is played  by }  the { LF reverse}
projection operator that allows the full reconstruction of the
four-dimensional BS amplitudes, without loosing any physics
content~\cite{sales00,sales01,adnei07}. The price to be paid,
however, is a rather complicated effective three-dimensional
interaction, that is derived from the full four-dimensional one. In
particular, the effective three-dimensional interaction is obtained
from the proper projection of the quasi-potential, expanded in
powers of the four-dimensional interaction { present in the BS
equation}. { In conclusion},
 such an approach is useful if   the convergence rate of the expansion allows
to adopt a truncated series, at a nontrivial low order, in the
actual calculations. The procedure of LF projection and
expansion/truncation of the three-dimensional interaction is
discussed in Ref.~\cite{sales00} for the two-boson case and in
Ref.~\cite{sales01} for the two-fermion case.

By using the two previously introduced ingredients, namely QP
approach and LF projection, we  first investigate the possibility to
express the matrix elements of the four-dimensional em current
operator, for an interacting two-fermion system, through matrix
elements of a three-dimensional LF current operator between valence
states,  without introducing any truncation (see \cite{karmanovnpb}
for a formal definition of the valence state). In particular, it
will be shown that the three-dimensional LF current operator fulfills
WTI.  This initial analysis   will allow
us to define LF charge operators and the form of the WTI onto the LF
hyperplane. After introducing a proper truncation procedure of the
LF current operator, it will be presented a systematical way for
constructing workable approximations of the matrix elements of the
LF current by increasing the order of truncation of the QP
expansion. It will be shown that the truncated LF current fulfills
WTI with the LF charge operators obtained in the non truncated case,
leaving a complete study of the relativistic covariance under
dynamical transformations to a future investigation. In order to
illustrate our procedure in an actual case,  it will be analyzed
the lowest non trivial truncation of a Yukawa model with chargeless
boson exchange, in ladder approximation.

As well-known (see, e.g. Ref. \cite{brodsky})  a peculiar feature of
any LF description of fermionic systems, is the so-called
instantaneous (in LF time) propagation. The treatment of this issue
makes sharply different the study of the em current for fermionic
systems from the bosonic case, already analyzed in detail in
Ref.~\cite{adnei07}, and in what follows this point will be often
emphasized.

In general, the construction of a conserved current operator acting
on the valence component of the LF wave function is a challenging
problem. For two-boson systems Kvinikhidze and
Blankleider~\cite{kvi03} developed gauging techniques for solving
this problem. They were able to obtain a current that satisfies a
WTI and therefore is conserved. It is remarkable that for the
two-boson case the current operator derived in Ref.~\cite{adnei07}
is equivalent to the one obtained with gauging
techniques~\cite{kvi03}. An earlier perturbative approach to the WTI
within LF quantization was performed in Ref.~\cite{naus98}.

This work is organized as follows. In Sect. II, the four-dimensional
expression of the em current operator and the associated WTI in the
context of BS formalism, that leads to current conservation, is
briefly recalled. In Sect. III, we illustrate the LF time projection
technique, that allows one to eliminate relative LF time, within a
QP approach applied to the BS equation for two-fermion systems. In
Sect. IV, it is shown the connection of the two-fermion valence LF
wave function, for bound and scattering states, with the
corresponding BS amplitude. In Sect. V, the three-dimensional
current operator is introduced and the fulfilled WTI is discussed.
In Sect. VI, we propose a
truncated form of the em current operator that satisfies the WTI
at any given order, and we explicitly show that the current
conservation is fulfilled by the matrix elements calculated between
valence states. In Sect. VII, we present the formal construction
in an actual case: a Yukawa model with a chargeless boson
exchange, in  ladder approximation. Our conclusions are drawn in
Sect. VIII.

\section{Four-dimensional EM Current Operator}

The general field-theoretical description of an interacting
two-fermion system is given by the BS  equation, where the driving
term, the interaction $V(K)$, is assumed to be based on the
irreducible exchange of bosons. In what follows self-energy
corrections will be omitted and considered elsewhere. Let us remind
that the total four-momentum $K$ is conserved, and all BS operators,
as $V(K)$, depend parametrically on $K$ { (see, e.g.
\cite{adnei07})}. As well-known, the interaction $V(K)$ yields the
transition matrix $T(K)$, i.e., \be T(K)=V(K)+VG_0(K)T(K), \label{tls}
\ee
 and the full Green's function, $G(K)$, i.e.,
\be
G(K)=G_0(K)+G_0(K)V(K)G(K)=G_0(K)+G_0(K)T(K)G_0(K)~ ,\label{gfull}
\ee
where $G_0(K)$ is the free Green's function of two fermions, with
four-momentum operators $\widehat k_j$ (j=1,2), viz.,
\be
G_0(K)={\imath^2 \over 2\pi} ~\frac{ \psla \widehat {k}_1
+m_1}{\widehat k_1^2-m_1^2+i\varepsilon}~ \frac{ \psla \widehat
k_2+m_2}{\widehat k_2^2-m_2^2+i\varepsilon}~,\label{livre}
\ee
and $\widehat k_2=K-\widehat k_1$. For convenience a factor of
$1/2\pi$ has been included in the above definition.

The BS amplitude $| \Psi \rangle$, {  dependent upon internal
variables}, satisfies (the $\lim_{\varepsilon\to 0}$ is understood)
\be G^{-1}(K)\left| \Psi \right\rangle =0~, \label{bseqinv} \ee with
appropriate boundary conditions for bound and scattering
states~\cite{itz}.

The em current operator, ${\mathcal J}^\mu$, corresponding to the
interaction $V(K)$ has a free term, ${\mathcal J}_0^\mu$, and
another one, ${\mathcal J}_I^\mu$, which depends on the interaction,
as dictated by the commutation rules between ${\mathcal J}^\mu$ and
the generators of the Poincar\`e group, i.e., \be
\mathcal{J}^\mu(Q)=\mathcal{J}^\mu_0(Q)+\mathcal{J}^\mu_I(Q).\label{4dcurr}
\ee where { ${\mathcal J}_I^\mu$ depends parametrically on the
4-momentum transfer $Q=K_{f}-K_{i}$.} { In particular the Lorentz
covariance of ${\mathcal J}^\mu$ imposes that (see, e.g.
\cite{LPS98}) \be \Lambda^\mu_\nu~
\mathcal{J}^\nu(Q')=D^{J_f}\left[W(\Lambda^{-1},K_f)\right]^{-1}~
\mathcal{J}^\mu(Q)~D^{J_i}\left[W(\Lambda^{-1},K_i)\right]\label{lcov}\ee
where $\Lambda$ is a Lorentz transformation, ${Q'}=\Lambda^{-1}~Q$,
 $J_{i(f)}$
is the total spin of the initial (final) system,  and $D^J$ is the
unitary representation of
the Wigner
rotation, $W(\Lambda^{-1},K)$, corresponding to $\Lambda$. As to the BS amplitude one has
\be
\Psi^\prime_{JM_J}(k'_1,K')=\sum_{M'_J}~ S^{-1}_1(\Lambda)~ S^{-1}_2(\Lambda)
D^J_{M'_J M_J}\left[W(\Lambda^{-1},K)\right]~ \Psi_{JM'_J}(k_1,K)
\ee
where $k'_1=\Lambda^{-1}~ k_1$, ${K'}=\Lambda^{-1}~K$,
and $S_{1(2)}(\Lambda)$ is the spinorial representation of the
transformation $\Lambda$ \cite{Primack69}.}

Notably, the four-current
 satisfies the following Ward-Takahashi
identity~ (see, e.g., \cite{itz,gross}) \be Q_\mu{\mathcal
J}^\mu(Q)=G^{-1}(K_f)\widehat e~-\widehat e~G^{-1}(K_i)
, \label{jwti} \ee where $\widehat e=\widehat e_1+\widehat e_2$ and  $\widehat e_j$ is the
charge operator for the fermion $j$, with matrix elements given by
\be \langle k_j|\widehat
e_j|p_j\rangle=e_j\delta^4\left(k_j-p_j-Q\right). \label{charge} \ee
{ Since the inverse of the full interacting Green's function is
given by \be G^{-1}(K)=G^{-1}_0(K) -V(K), \label{gminus1} \ee and \be
Q_\mu{\mathcal J}_0^\mu(Q)= G^{-1}_0(K_f)\widehat e~-\widehat
e~G^{-1}_0(K_i)  \label{freewti}\ee then the
interacting current fulfills the following relation \be
Q_\mu{\mathcal J}_I^\mu(Q)= \widehat e~ V(K_i)-V(K_f)\widehat e~
.
 \label{jwti0}
\ee}
Once we consider the matrix elements $\langle \Psi_{f}|{\mathcal
J}^\mu(Q)\left| \Psi_{i}\right\rangle$, current
conservation can be explicitly obtained through Eq.~(\ref{bseqinv}),
i.e., \be Q_\mu\left\langle \Psi_{f}\right|{\mathcal J}^\mu(Q)\left|
\Psi_{i}\right\rangle=\langle \Psi_{f}|~\left[G^{-1}(K_f) \widehat
e~-\widehat e~G^{-1}(K_i)\right]~|\Psi_{i}\rangle =0.\ee

\section{LF-time Projection and the Quasi-Potential Approach for Fermions}

Following Ref. \cite{sales01}, where the   LF projection of the BS
equation for a fermionic system  was investigated, here we briefly resume
 the QP
formalism for the LF projection of the two-fermion BS equation (see
also Refs.~\cite{prlwilson,prdwilson,carbonell}), { and we introduce
a more compact and efficient operator notation}.

Crucial for the LF
projection is the separation of the fermion propagator in an
on-shell term and in an instantaneous one, viz.
\begin{eqnarray}
\frac{\psla{k}+m}{k^2-m^2+i\varepsilon}=
\frac{\psla{k}_{on}+m}{k^+(k^--k^-_{on} +{i\varepsilon\over
k^+})}+\frac{\gamma^+}{2 k^+} \ , \label{instant}
\end{eqnarray}
where $k^-_{on}=(\vec k_{\perp}^2+m^2)/k^+$ is the on-minus-shell
momentum. The second term in Eq. (\ref{instant}) does not lead to a
free propagation in the global time, since the Fourier transform is
divergent and yields $\delta(x^+)$, i.e. an instantaneous  (in the LF
time) propagation. This term makes the treatment of a fermionic
system basically different from the treatment of a bosonic one. The LF projection
is based on the on-shell part of the two-fermion free propagator
of Eq.~(\ref{livre}), i.e.
\begin{eqnarray}
\overline{G}_{0}(K)  := \frac{  i^2}{2\pi } \frac{\left(\psla
\widehat k_{1on}+m_{1}\right) \left(  {\psla \widehat
k}_{2on}+m_{2}\right)
}{\widehat{k}_{1}^{+}(K^{+}-\widehat{k}_{1}^{+})\left( \widehat
k_{1}^{-}-\frac{\widehat{\vec{k}}_{1\perp }^{2}+m_{1}^{2}
-i\varepsilon}{\widehat{k}%
_{1}^{+}}\right) \left( K^{-}-\widehat
k_{1}^{-}-\frac{\widehat{\vec{k}}_{2\perp
}^{2}+m_{2}^{2}-i\varepsilon}{K^{+}-\widehat{k}_{1}^{+}}\right) } \
. \label{livon}
\end{eqnarray}

{ The role of $\overline{G}_{0}(K)$ will become more and more clear
as} the LF integral equations for the transition matrix, the valence
Green's function, the valence wave function and etc., { will be}
derived. { It should be pointed out that} the instantaneous terms
are fully restored into the theory through the three-dimensional
effective interaction { (see below  and also Sect. IV)}. Once the
full structure of the propagator is taken into account, the matrix
elements of the three-dimensional em current operator become equal
to the corresponding four-dimensional quantities, as will be
discussed in Sects. IV and V.

The free Green's functions $G_{0}(K)$ and $\overline G_{0}(K)$ are
four-dimensional operators which depend upon the four-momenta of the
two fermions. Let us introduce the LF free Green's function,
$g_{0}(K)$, that is a three-dimensional operator  which depends upon
the LF momenta $(k^+_i,\vec k_{i\perp })$ only. It is obtained from
the four-dimensional on-shell Green's function by projection, i.e.
\be g_{0}(K)=~|\overline{G}_{0}(K)|:=  \int dk_{1}^{\prime
-}dk_{1}^{-}\left\langle k_{1}^{\prime -}| \overline{G}_{0}(K)|
k_{1}^{-}\right\rangle \ \label{2.11a}\\&& =i\theta
(K^{+}-\widehat{k}_{1}^{+})\theta (\widehat{k}_{1}^{+})\frac{
2m_{1}2m_{2}\Lambda _{+}(\widehat{k}_{1on})\Lambda _{+}(\widehat{k}
_{2on})}{\widehat{k}_{1}^{+}(K^{+}-\widehat{k}_{1}^{+})\left(
K^{-}-\widehat{
k}_{1on}^{-}-\widehat{k}_{2on}^{-}+i\varepsilon\right) } \ ,
\label{2.11b} \ee where one can choose $K^{+}\ >\ 0$ without any
loss of generality, and $\Lambda_{+}(\widehat{k}_{on}) =\left(
\widehat{\rlap\slash k}_{on}+m\right) /2m$ is the positive energy
spinor projector. The vertical bar, $|$, on the right (left)
indicates that the minus component present in the ket, $|k^-\rangle$
(in the bra, $\langle k^-|$) is integrated out, namely, through the
vertical bar operation one projects on the LF hypersurface $x^+=0$ {
(see Refs. \cite{sales00,sales01,adnei07})}. Thus, in this paper the
two vertical bars in Eq.~(\ref{2.11a}) {\it do not mean} the
absolute value; they indicate, indeed, the transition from a
four-dimensional operator to a three-dimensional one. It is worth
noting that { $g_0(K)$ yields the free global propagation of two
on-shell fermions, and its} inverse  exists in the valence sector,
since $\Lambda _{+}(\widehat{k}_{1on})\Lambda
_{+}(\widehat{k}_{2on})$ is the representation of the identity in a
two-particle spinor space. { Furthermore, one can express   the
three-dimensional $g^{-1}_0(K)$ in terms of the  corresponding
four-dimensional quantity, $G^{-1}_0(K)$, as follows (see Appendix
A, Eq. (\ref{Ag0})) \be
g^{-1}_0(K)=\overline\Pi_0(K)~G^{-1}_0(K)~\Pi_0(K) \label{g0m1}\ee
where we define the {\em free reverse LF projection operator} (cf the following
Eq. (\ref{psi1})) and
its LF conjugated, as \be \Pi _0(K):=\overline G_0(K) |~g^{-1}_0(K)
\quad \quad \overline\Pi_0(K):=g^{-1}_0(K) ~| \overline G_0(K) .
\label{freeproj}\ee An explicit expression of $\Pi _0(K)$ can be
found in the Appendix of Ref. \cite{sales01}. These operators have
the following properties, as shown by the definitions in Eq.
(\ref{freeproj}) and from Eq. (\ref{Ag0}), \be |\Pi_0(K)={\text
I}\hspace{4.35cm}\overline\Pi_0(K)|={\text I} \nonu | \overline
G_0(K)~ G^{-1}_0(K)~\Pi_0(K)= {\text I}\quad \quad \quad
\overline\Pi_0(K)~ G^{-1}_0(K)~\overline G_0(K)|= {\text I}
\label{pipro}\ee where ${\text I}$ is the identity in the LF
three-dimensional space.

 { The LF free state, $|\phi_0\rangle$,  is a solution of}
\be g^{-1}_0(K)~|\phi_0\rangle =0 \label{freeLF}\ee then from Eq.
(\ref{g0m1}) one has\be
\overline\Pi_0(K)~G^{-1}_0(K)~\Pi_0(K)~|\phi_0\rangle =0
\label{phi0}\ee The last equation leads to the following
transformation property between the non-interacting BS amplitude and
the corresponding LF three-dimensional state, viz \be
|\Psi_0\rangle=\Pi_0(K)~|\phi_0\rangle\label{psi1}\ee
Indeed, one obtains
the eigenequation for $|\Psi_0\rangle$ by applying $G^{-1}_0(K)$ to
Eq. (\ref{psi1}) { and performing the understood $\lim_{\varepsilon \to 0}$.
Then, on the rhs of Eq. (\ref{psi1}) one can use
  Eqs. (\ref{freeproj}) and (\ref{I1})
(that holds for any $\varepsilon$) and can shift $\lim_{\varepsilon \to 0}$
to $g^{-1}_0(K)~|\phi_0\rangle$}, namely one has
\be
G^{-1}_0(K)~|\Psi_0\rangle=0\label{freeBS}\ee
Furthermore, one gets
two expressions of the inverse of Eq. (\ref{psi1}) by using Eqs. (\ref{pipro}) and
(\ref{psi1}), viz
\be ||\Psi_0\rangle= |\phi_0\rangle \ee
and
\be
 |\overline G_0(K)~ G^{-1}_0(K)~|\Psi_0\rangle=
|\phi_0\rangle\label{BSLF}\ee}
{ In Eq. (\ref{BSLF}) the understood  $\lim_{\varepsilon \to 0}$
affects the whole
lhs.} Notably, from the uniqueness of the
solutions of Eq. (\ref{freeBS}) and applying Appendix A of Ref.
\cite {sales01}, one can demonstrate that for any given solution
$|\Psi_0\rangle$ of Eq. (\ref{freeBS}) one has a LF wave function
obtained by Eq. (\ref{BSLF}), that in turn fulfills the
eigenequation (\ref{freeLF}) and yields the initial $|\Psi_0\rangle$
through Eq. (\ref{psi1}). Namely, there is a one-to-one relation
between a given four-dimensional BS amplitude $|\Psi_0\rangle$ and
the corresponding three-dimensional $|\phi_0\rangle$. The
non-interacting operators $\Pi_0(K)$ and $\overline\Pi_0(K)$ connect
three- and four-dimensional quantities. In the next Section the
corresponding interacting operators will be given.

The QP formalism makes use of the { four-dimensional} auxiliary
Green's function $\widetilde G_0(K)$ defined by
\be
\widetilde G_0(K):= \overline G_0(K)|~g_0^{-1}(K)~| \overline G_0(K)=
\Pi _0(K) g_0(K) \overline\Pi_0(K)=\nonu = \overline
G_0(K)|~ \overline\Pi_0(K)= \Pi _0(K)~| \overline G_0(K)    ~,
\label{g0tilde}
\ee
{ This operator has the following useful properties \be
 |\widetilde G_0(K)=|\overline G_0(K) \quad \quad \quad \widetilde
 G_0(K)|=\overline G_0(K)|~~.
\ee } {  Let us apply the QP formalism to the four-dimensional}
transition matrix (see Refs. \cite{wolja,sales00,sales01}), i.e.
\begin{eqnarray}
T(K)=W(K)+W(K)\widetilde{G}_{0}(K)T(K)=W(K)+T(K)\widetilde{G}_{0}(K)W(K),
\label{2.1}
\end{eqnarray}
where  the effective interaction $W(K)$,  in turn,
is a solution of
\begin{equation}
W(K)=V(K)+V(K)\Delta_0(K)W(K)\ , \label{2.3a}
\end{equation}
with \be \Delta_0(K):=G_{0}(K)-\widetilde{G}_{0}(K) ~.
\label{delta0}  \ee { The four-dimensional quantity $\Delta_0(K)$
clearly contains the instantaneous terms and has the following
properties \be \gamma^+_1~\gamma^+_2~\Delta_0(K)|=0 \quad \quad
\quad \gamma^+_1~\gamma^+_2~|\Delta_0(K)=0 \label{deltapp}\ee It is
worth noting that in Eq.~(\ref{2.1}) the driving term $V(K)$ is
substituted by $W(K)$, in order to increase the rate of convergence
of the iterative solution of the transition matrix (for numerical
studies of the convergence see, e.g., \cite{sales00,miller3}). In a
similar manner,} Eq.~(\ref{2.3a}) can be solved by iteration,
obtaining \be W(K)=\sum_{i=1}^\infty W_i(K), \label{sumw} \ee with
$W_i(K)=V(K)\left[\Delta_0(K)V(K)\right]^{i-1}$. The physical meaning of the
convergence of this series will be discussed in Sect. VI.

{ The three-dimensional LF transition matrix can be introduced
through (see also \cite{sales01})
 \be t(K)=\overline\Pi_0(K)~T(K)~\Pi _0(K)
 \label{3.1}
 \ee
Then, by using Eq. (\ref{2.1}), one has
\be t(K)=\overline\Pi_0(K)~
 \left[W(K)+W(K)\widetilde{G}_{0}(K)T(K) \right]~\Pi _0(K)=\nonu
 =w(K)+w(K)g_{0}(K)t(K)=w(K)+w(K)g(K)w(K)\ ,  \label{3.3}
\ee
  where
 the three-dimensional driving term, $w(K)$ is obtained from the four-dimensional
interaction $W(K)$  according to
\be
w(K):= {\overline \Pi}_{0}(K)W(K) \Pi_{0}(K) \ .  \label{3.4}
\ee
and  the three-dimensional
 interacting Green's function is given by
 \be
 g(K)=g_0(K)+g_0(K)t(K)g_0(K) .
  \label{LFRESOLV0}
 \ee

Note that i) the positions of $\overline\Pi_0(K)$ and  $\Pi_0(K)$
 in Eq. (\ref{3.1}),
leading to the integrations over the external  minus components, straightforwardly
 indicate that $t(k)$ is a three-dimensional quantity (for comparison see Eq.
 (\ref{g0tilde}), where the four-dimensional quantity $\widetilde G_0(K)$ is
 defined, and the integrations are on the internal minus components);
 ii) the effective interaction $w(K)$ contains the coupling of the
valence sector to the higher Fock-state components of the wave
function through $W(K)$~\cite{hierareq}.}

{ Finally, the four-dimensional $T(K)$, can be rewritten as follows
\be T(K)={W}(K)+W(K)~\Pi_0(K)~ g(K)~\overline \Pi_0(K)~W(K)
\label{3.2} \ee Such an expression is useful to discuss the relation
between the four-dimensional BS amplitude for a bound state and the
corresponding three-dimensional valence wave function (cf Sect. IV).
}

The LF interacting Green's function, or resolvent, can also be
written as a function of $w(K)$ as follows: \be
g(K)=g_0(K)+g_0(K)w(K)g(K)=g_0(K)+g(K)w(K)g_0(K) . \label{LFRESOLV}
\ee It should be pointed out that $g(K)$ is the Fourier transform of
the global propagator of two interacting fermions  between two LF
hypersurfaces with given $x^+$'s. { Its inverse can be easily
related to the four dimensional $ G^{-1}(K)$ by using Eqs.
(\ref{g0m1}), (\ref{3.4}) and (\ref{2.3a}). As a matter of fact, one
has \be g^{-1}(K )=g^{-1}_0(K )-w(K)={\overline \Pi}_{0}(K)
 \left [G^{-1}_0(K)-W(K) \right ] \Pi_{0}(K)\nonu={\overline \Pi}_{0}(K)
 \left [G^{-1}_0(K)-V(K)\left (1 +\Delta_0(K)~W(K)\right ) \right ]
 \Pi_{0}(K)\nonu={\overline \Pi}_{0}(K)
 G^{-1}(K)~\left [1 +\Delta_0(K)~W(K)\right ] \Pi_{0}(K)
-{\overline \Pi}_{0}(K)G^{-1}_0(K)\Delta_0(K)W(K) \Pi_{0}(K)\ee} {
The last term is vanishing, as can be shown by using Eqs.
(\ref{freeproj}),
 (\ref{I2}) and the second relation in Eq. (\ref{deltapp}). Finally one obtains
 \be  g^{-1}(K )={\overline \Pi}_{0}(K)~G^{-1}(K) ~\Pi(K) \label{gfm1}\ee
 where
 \be \Pi(K)=~\left [1 +\Delta_0(K)~W(K)~\right ]~ \Pi_{0}(K)\label{invop}\ee is the
 {\em interacting LF
 reverse projection operator}.} { From analogous steps one can obtain
 the corresponding  LF conjugated operator, given by
 \be \overline \Pi(K)= ~\overline\Pi_{0}(K)~
 \left [1 +W(K)~\Delta_0(K)~\right ]\label{invopb}\ee
 It will be very useful in what follows to note that by using Eqs. (\ref{I1}),
 (\ref{I2}), (\ref{pipro}) and (\ref{deltapp}) one
 has in the three-dimensional space
 \be  |\overline G_0(K) ~G^{-1}_0(K)~\Pi(K)= ~{\rm I}
 \quad \quad \quad \overline\Pi(K)
 ~G^{-1}_0(K)~\overline G_0(K)| =~{\rm I}
 \label{invpi}\ee
 }
  The solution of the
following three-dimensional equation,
\begin{eqnarray}
g^{-1}(K )\left| \phi \right\rangle = \left[g^{-1}_0(K
)-w(K)\right]~\left| \phi \right\rangle=0 \label{valeneq}
\end{eqnarray}
with appropriate boundary conditions for bound and scattering
states, is the valence component of the LF wave
function~\cite{brodsky,karmaprep}. It turns out that the full
complexity of the Fock-space comes through the effective
interaction. { However,} the truncation of the quasi-potential (see
Eq. (\ref{sumw})) limits the number of Fock-components involved in
the construction of effective interaction to be used to obtain the
valence wave function~\cite{hierareq}.  Then, one could argue that
the convergence rate of the QP expansion is related to the smallness
of the probability for the higher Fock-components.

{ Inserting Eq. (\ref{gfm1}) in the eigenequation (\ref{valeneq})
one has \be g^{-1}(K )\left| \phi \right\rangle ={\overline
\Pi}_{0}(K)~G^{-1}(K) ~\Pi(K)\left| \phi \right\rangle =0\ee that
leads to the following relation between the three-dimensional
valence component and the four-dimensional BS amplitude \be~\left|
\Psi \right\rangle=\Pi(K)~\left| \phi
\right\rangle\label{psitolf}\ee Note that $| \Psi \rangle$ fulfills
Eq. (\ref{bseqinv}) (as can be seen by using the results in Appendix
A of \cite{sales01}). Furthermore, by applying Eq. (\ref{invpi}),
one gets \be |\overline G_0(K) ~G^{-1}_0(K)~\left| \Psi
\right\rangle=~ \left| \phi \right\rangle\label{invpsi}\ee In the
next Section more details will be given about the one-to-one
relation between the four-dimensional $| \Psi \rangle$ and the
three-dimensional $| \phi \rangle$.}

Finally, let us remind that,  the on-mass-shell matrix elements of $T(K)$,
which define the two-fermion scattering amplitude are identical to
the ones obtained from $t(K)$~\cite{sales01}.

\section{ The LF valence wave function and the BS amplitude }

{ In this Section, we will analyze the interacting operator
$\Pi(K)$, Eq. (\ref{invop}), that generates the full BS amplitude,
for bound and scattering states of two-fermion systems, starting
from the corresponding valence wave functions. It should be
emphasized the key role played by $\Pi(K)$, when the matrix elements
of a four-dimensional operator acting on the BS amplitudes are
considered. In particular, the LF reverse projection allows  to
express those matrix elements in terms of matrix elements of
effective operators acting on the valence wave functions.}

The relation between the BS amplitude for a two-fermion bound
system, $\left|\Psi_B\right\rangle$, and the corresponding valence
wave function has been derived in~\cite{sales01}, and reads as
\begin{eqnarray}
\left| \Psi _{B}\right\rangle &=&G_{0}(K_{B})W(K_{B})~\Pi_0(K_B)
~\left|\phi _{B}\right\rangle \ . \label{3.6a}
\end{eqnarray}
This can be obtained through the analysis of the poles  of
Eq.~(\ref{3.2}).

 Since the valence wave function is the
solution of
\begin{eqnarray}
\left| \phi _{B}\right\rangle &=&g_{0} (K_{B})w(K_{B})\left| \phi
_{B}\right\rangle \ , \label{lfbse}
\end{eqnarray}
the vanishing quantity $\overline G_0(K_B)|~\left(g^{-1}_{0}
(K_{B})-w(K_B)\right)\left| \phi _{B}\right\rangle =0$ can be added
to Eq.~(\ref{3.6a}) in order to { recover the LF reverse projection
as given in Eq. (\ref{invop}), i.e.}
\begin{eqnarray}
\left| \Psi _{B}\right\rangle
&=&\left[1+\Delta_0(K_B)W(K_{B})\right] {\Pi}_{0}(K_{B})~\left| \phi
_{B}\right\rangle =\Pi(K_B)~\left| \phi
_{B}\right\rangle \ . \label{psicov}
\end{eqnarray}
{ Therefore any $\left| \Psi _{B}\right\rangle$ can be generated
through ${\Pi}(K_B)$ from the corresponding valence wave function
$\left| \phi _{B}\right\rangle $, eigensolution of Eq.
(\ref{valeneq}). Moreover, from Eq. (\ref{invpsi})
 one  has
\be |\overline G_0(K_B) ~G^{-1}_0(K_B)~\left| \Psi_B \right\rangle=~
\left| \phi_B \right\rangle\ee} { Let us now show that the operator
${\Pi}(K)$ connects BS amplitudes and valence wave functions
 for scattering
states, as well.

The BS amplitude for scattering states satisfies the four-dimensional
Lippman-Schwinger type inhomogeneous equation:
\be
|\Psi^+\rangle=|{\Psi}_0\rangle+G_0(K)T(K)|{\Psi}_0\rangle
=\left[1+G_0(K)T(K)\right]{\Pi}_0|\phi_0\rangle ~ , \label{psi+}
\ee
where we have made use of Eq. (\ref{psi1}), that univocally relates
$|{\Psi}_0\rangle$ to $|\phi_0\rangle$. In the three-dimensional space the
scattering solution of Eq. (\ref{valeneq}) is given by
\be
 |\phi^+\rangle=\left[1+g(K)w(K)\right]|\phi_0\rangle
\nonu
=\left[g_0(K)+g(K)w(K)g_0(K)\right]g^{-1}_0(K)|\phi_0\rangle
=g(K)g_0^{-1}(K)|\phi_0\rangle
 ~,\label{phi+}
\ee
which implies   the formal identity
$g_0(K)g^{-1}(K)|\phi^+\rangle=|\phi_0\rangle$. Then, by using Eq.
(\ref{rev1}), one gets
\be
|\Psi^+\rangle
=\left[1+G_0(K)T(K)\right]{\Pi}_0(K)|\phi_0\rangle=\nonu
\left[1+G_0(K)T(K)\right]{\Pi}_0(K)~g_0(K)~g^{-1}(K)~|\phi^+\rangle=\nonu
=\Pi(K)|\phi^+\rangle ~ . \label{psi+1}
\ee
}
From Eq.
(\ref{invpsi}), the valence component of the LF wave function can be
obtained directly from the BS amplitude, i.e. \be |\overline G_0(K)
G_0^{-1}(K)~\left|\Psi^+\right\rangle=|\phi^+\rangle\ee { It is
worth noting that the operator $|\overline G_0(K) G_0^{-1}(K)$
  cuts the instantaneous terms and projects onto the LF
hyperplane (through the $k^-$ integration), while ${\Pi}(K)$ reconstructs the
full structure of the four-dimensional BS amplitude through the instantaneous
terms contained in   $G_0$ (see $\Delta_0$) and  $W$. Moreover, the
 instantaneous terms affect
 the effective interaction $w$
that determines $\left| \phi _{B}\right\rangle$ and  $\left|
\phi^+\right\rangle$ (cf
Eq.~(\ref{valeneq})).}

We should point out that the normalization of the BS amplitude for a
bound state~\cite{LMT65} expressed in an operatorial form,
${\mathcal N}(K)$, can be  mapped onto an expectation value of a
proper three-dimensional operator viz., $\langle \Psi |{\mathcal
N}(K)|\Psi\rangle=\langle \phi |\overline\Pi(K){\mathcal
N}(K)\Pi(K)|\phi\rangle$. Note that on one side, the BS amplitude
normalization corresponds to the sum over the probabilities of each
Fock component in the full LF wave function~\cite{karmanovnpb}, and
on the other side the full complexity of the Fock-space is
summarized in the three-dimensional operator $\overline\Pi(K){\mathcal
N}(K)\Pi(K)$.

\section{LF Ward-Takahashi Identity}

Once we have the relation between the 4-dimensional BS amplitude and
the 3-dimensional LF valence wave function (see Eq. (\ref{psitolf})),
the LF em current operator can be obtained from the matrix
element of the four-dimensional current. As a matter of fact, one
has for both scattering and bound states
\begin{eqnarray}
\left \langle\Psi_{f}\right|{\mathcal J}^\mu(Q)\left|
\Psi_{i}\right\rangle=\langle\phi_{f} |j^\mu(K_f,K_i) |\phi_i\rangle
~,\label{lfc1}
\end{eqnarray}
where  the  three-dimensional LF current operator, acting on
the valence wave functions, is defined as follows
\begin{eqnarray}
j^\mu(K_f,K_i) :=  \overline \Pi(K_f){\mathcal J}^\mu(Q)\Pi(K_i)
~.\label{lfc2}
\end{eqnarray}
Using Eqs. (\ref{invop}) and (\ref{invopb}), one can put in evidence the
dependence of the LF current operator upon the effective interaction $W(K)$, viz
\begin{eqnarray}
j^\mu(K_f,K_i) := \overline\Pi_0(K_f)\left[1+
W(K_f)\Delta_0(K_f)\right]{\mathcal
J}^\mu(Q)\left[1+\Delta_0(K_i)W(K_i)\right]
\Pi_{0}(K_i)~.\label{lfc2L}
\end{eqnarray}
It is worth noting that Eq.~(\ref{lfc2L}) generalizes to scattering
states the expression for the three-dimensional current already
derived in Ref. \cite{sales01} for bound states. It should be
pointed out that ${\mathcal J}_\mu$ is  Poincar\'e covariant (cf Eq.
(\ref{lcov}) for the Lorentz covariance). Since all the matrix
elements of the lhs are properly related through the  Lorentz
transformations, the investigation of the covariance properties of
the operator $j_\mu$, within the full theory, does not represent a
stringent question in view of the equality in Eq. (\ref{lfc1}). But,
given the following  development of the truncated approach, where
the full covariance is broken, it is important to investigate the
transformation properties of $j_\mu$. As discussed elsewhere
\cite{MFPSSup},  we can anticipate that the covariance with respect
to the kinematical transformations can be demonstrated after
introducing  new factors in  the vertical bar operation (cf also the
factor $\widehat \Omega$ in  Ref. \cite{hierareq}).

In what follows, our task  is to find a suitable definition of the
LF charge operator that allows to write the four-divergence of   the
LF current operator in terms of the inverse of the Green's functions
$g^{-1}(K_f)$ and $g^{-1}(K_i)$. { Such an investigation  is
fundamental to obtain the WTI for the truncated  LF current (see the
next Section).    } It should be pointed out that a similar analysis
has been already performed for bosonic systems in \cite{adnei07}.

In order to find the LF charge operator, the four-dimensional
divergence of the LF
current can be written as follows by using
  Eqs. (\ref{lfc2}) and
(\ref{jwti})
\be
Q_\mu j^\mu(K_f,K_i) = \overline \Pi(K_f)\left[G^{-1}(K_f)\widehat
e-\widehat eG^{-1}(K_i)\right]\Pi(K_i) ~,\label{lfc3.1}
\ee
with $\widehat e=\widehat e_1+\widehat e_2$.

{ By applying Eqs. (\ref{rev2}) and (\ref{rev3}) of Appendix B, one
gets \be Q_\mu j^\mu(K_f,K_i) =  g^{-1}(K_f)~~|{\overline
G}_0(K_f)~G^{-1}_0(K_f)~\widehat e~ \Pi(K_i) - \overline
\Pi(K_f)~\widehat e~G^{-1}_0(K_i)~{\overline G}_0(K_i)|~g^{-1}(K_i)
=\nonu= g^{-1}(K_f)~\widehat {\cal Q}^L_{LF}- \widehat {\cal
Q}^R_{LF}~ g^{-1}(K_i) \label{lfc11}  \ee where  {\it left} and {\it
right} LF charge operators have been introduced. Such operators are
defined as follows \be \widehat {\cal Q}^L_{LF}=|{\overline
G}_0(K_f)~G^{-1}_0(K_f)~\widehat e~ \Pi(K_i)= |{\overline
G}_0(K_f)~G^{-1}_0(K_f)~\widehat e~ \Pi_0(K_i) \label{eilf} \ee \be
\widehat {\cal Q}^R_{LF}~=\overline \Pi(K_f)~\widehat e
~G^{-1}_0(K_i)~{\overline G}_0(K_i)| =\overline \Pi_0(K_f)~\widehat
e ~G^{-1}_0(K_i)~{\overline G}_0(K_i)|
 \label{eflf}\ee
where Eqs. (\ref{I1}), (\ref{I2}) and (\ref{deltapp}) have been
used, as well as  the absence of a Dirac structure in the operator
$\widehat e$. It is very important to note that the LF charge
operators are non interacting operators acting on the two-fermion
spinor space. From Eqs. (\ref{I2}) and  (\ref{pro3}), we can  obtain
the explicit expression for the {\em  left} LF  charge operator for
particle $1$ \be \widehat {\mathcal Q}^L_{1LF}= \Lambda_+(\widehat
k_{1on})\frac{m_1}{\widehat k^{+}_1}\gamma_1^+~\widehat e_{1LF}
\Lambda_+(\widehat k_{1on})\Lambda_+(\widehat k_{2on})
~.\label{lfc9} \ee where { the notation $\widehat e_{1LF}$ indicates
the three-dimensional LF counterpart of the operator $\hat e$,
 Eq.~(\ref{charge}), with matrix elements given by}
\begin{eqnarray}
\langle k^{\prime+}_1,{\vec k}^\prime_{1\perp}|\widehat
e_{1LF}|k_1^+,\vec
k_{1j\perp}\rangle:=e_1\delta\left(k^{\prime+}_1-k^+_1-Q^+\right)~
\delta^2\left({\vec
k}^\prime_{1\perp}-\vec k_{1\perp}-\vec Q_{\perp}\right), \label{d3}
\end{eqnarray}

The corresponding  {\em right} operator
is
\be
\widehat {\mathcal Q}^R_{1LF}= \Lambda_+(\widehat k_{1on})~\widehat
e_{1LF}\frac{m_1}{\widehat k^{+}_1}\gamma_1^+  \Lambda_+(\widehat
k_{1on})\Lambda_+(\widehat k_{2on}) ~.\label{lfc10}
\ee
The operator $\gamma^+ m/k^+$ when sandwiched between LF spinors
gives the normalization condition. Let us remind that such an operator is
 the one
particle free charge operator of our model, since we do not include
fermion self-energy.} This also indicates the ingredients that ought
to be considered when  the full problem with self-energy insertions
is aimed. We will not discuss further this issue here, which is left
for a future study.

 From Eq. (\ref{lfc11}), current conservation straightforwardly follows by
taking the matrix elements between   three-dimensional interacting
states that are solutions of the wave equation (\ref{valeneq}) and
noting that the {\it left} and {\it right} charge operators do not
contain any $\imath\varepsilon$ dependence.

By multiplying both the left and right hand sides of Eq.~(\ref{lfc11}) by
$g(K_f)$ and $g(K_i)$, respectively, one gets
\begin{eqnarray}
Q_\mu g(K_f)j^\mu(K_f,K_i)g(K_i) =  \widehat{\mathcal Q}^L_{LF}g(K_i)-
g(K_f)\widehat {\mathcal Q}^R_{LF}~,\label{lfc12}
\end{eqnarray}
which corresponds to the LF projection of the 5-point function with
instantaneous terms cut from the external fermion legs.

 In conclusion the three-dimensional LF current given by Eq.
(\ref{lfc2}), acts on the LF valence wave functions and  fulfills WTI
(see Eq. (\ref{lfc12})).

\section{Ward-Takahashi Identity for the Truncated
LF Current}

In this section,  to obtain a workable approximation for the LF
current operator    that still fulfills WTI,
 we will analyze the consequences of a proper truncation of the
QP expansion  in the formal solution (Eq. (\ref{sumw})) of Eq. (\ref{2.3a}). As already
anticipated in the previous Section, the LF charges and the general form for
the LF WTI (cf Eq. (\ref{lfc11})),
will represent our fundamental ingredients. To simplify  notations, we will not show explicitly the
parametric dependence on total momentum $K$ in the operators.
Therefore, it is understood that  the operators on the left of the
charge operator depend upon $K_f$, while the ones on the right
depend upon $K_i$.

A naive substitution of $W$ in the current operator (\ref{lfc2L}),
by the truncated quasi-potential expansion
\be
W^{(n)}= \sum_{i=1}^{n}W_i \quad \quad {\rm with} \quad W_i=V[\Delta_0V]^{i-1}=
[V\Delta_0]^{i-1}V\label{Wn}\ee
namely by the truncated iterative solution
of Eq.~(\ref{2.3a}), does not lead to a  conserved three-dimensional
current. The same problem has been met and solved for two-boson
systems in Ref.~\cite{adnei07}. In this Section we will follow the
same procedure, but applied to the case of two-fermion systems.  In
the next Section an actual application for the Yukawa model will be
illustrated.

Let us first consider the conserved current, for the case $n=0$,
i.e. the current without  interaction. In this case one has
\be {{j}^{c(0)\mu}}:=\overline
\Pi_0 \mathcal{J}_0^\mu
 \Pi_0 =g_0^{-1}|\overline G_0{\mathcal J}^\mu_0\overline G_0|g_0^{-1}
 \ . \label{cvca0}
\ee { From Eqs. (\ref{cvca0}), (\ref{freewti}), (\ref{eilf}) and
(\ref{eflf}), WTI for $n=0$ reads: \be Q^\mu{j_\mu^{c (0)}}=
g_0^{-1}~|\overline G_0~G^{-1}_0 ~\widehat e ~\Pi_0-\overline \Pi_0~
\widehat e~G^{-1}_0~\overline G_0|~g_0^{-1}
=\nonu=g_0^{-1}\widehat{\mathcal Q}^L_{LF}-\widehat{\mathcal
Q}^R_{LF}g_0^{-1}. \label{divj0} \ee} The matrix elements of
Eq.~(\ref{divj0}) should be taken between solutions of
$g_0^{-1}|\phi_0\rangle=0$.

{ The matrix elements of $j_\mu^{c (0)}$ between free particle
states, as obtained explicitly in Appendix C  together with the
corresponding WTI, are  given by \be
 \langle k_1^{\prime +}\vec{k}_{1\perp }^{\prime
}\left|{{j}^{c(0)\mu}} \right| k_1^{+}\vec{k}_{1\perp }\rangle
= -i\theta\left(k^+_1\right)\theta\left(k_1^{\prime +}\right)
\theta\left(K^+_i-k^+_1\right) \theta\left(K^+_f-k^{\prime +}_1\right)
\nonu
\Lambda_+(k^\prime_{1on})\gamma_1^\mu\Lambda_+(k_{1on})\Lambda_+((K_i-k_1)_{on})
 \frac{K_i^{+}-k_1^{+}}{2m_2}\langle k_1^{\prime +},\vec
k^\prime_{1\perp}|\widehat e_{1,LF}|
k_1^{+}\vec{k}_{1\perp}\rangle
 +1\leftrightarrow 2  \label{cvme2}
\ee}

 From Eq. (\ref{lfc2L}) and cutting  at the first order  the
effective interaction (note that  $W^{(1)}=V$) one has \be
j^{(1)\mu}=j^{c(1)\mu}+ {\cal O}(V^2)+{\cal O}(V^3) \ee where the
first order contribution is given by \be j^{c(1)\mu}=
\overline{\Pi}_0\left[\mathcal{J}^\mu+
V\Delta_0\mathcal{J}^\mu_0+\mathcal{J}_0^\mu\Delta_0V\right]
{\Pi}_0=\nonu= j^{c(0)\mu}+\overline{\Pi}_0\left[\mathcal{J}^\mu_I+
V\Delta_0\mathcal{J}^\mu_0+\mathcal{J}_0^\mu\Delta_0V\right] {\Pi}_0
~, \label{jcn1} \ee { Only $j^{c(1)\mu}$ is a conserved current
operator in the corresponding valence sector.} Indeed, as shown in
details in Appendix D, one obtains the following WTI \be
Q^\mu{j_\mu^{c (1)}}={g_1}^{-1}~\widehat{\mathcal
Q}^L_{LF}-\widehat{\mathcal Q}^R_{LF}~{g_1}^{-1}~, \label{wtijcn1}
\ee where ${g_1}^{-1}={g_0}^{-1}-w^{(1)}$ and $w^{(1)}=
\overline{\Pi}_0~V~{\Pi}_0$.

 The matrix elements of
Eq.~(\ref{wtijcn1}) should be taken between solutions of
$g_1^{-1}|\phi_1\rangle=0$. { In  Sect. VII, we will discuss in
detail  the first-order LF current for the Yukawa model in ladder
approximation, and we will show that it is essential to include the
instantaneous terms, coming from $\Delta_0$ to obtain WTI for
$n=1$.}

 For the general case, where the interaction appears up to $n\geq 1$
times in the LF current operator, we  write
\be
j^{c (n)\mu}:=\overline{\Pi}_0\left[
\mathcal{J}^\mu+W_n\Delta_0\mathcal{J}^\mu_0+\mathcal{J}_0^\mu\Delta_0W_n+
\sum_{i=1}^{n-1}\left(W_i\Delta_0\mathcal{J}_0^\mu\Delta_0W_{n-i}
\right.\right. \nonu
\left.\left.+W_i\Delta_0\mathcal{J}^\mu+\mathcal{J}^\mu\Delta_0W_i\right)
+\sum_{j=2}^{n-1}\sum_{i=1}^{j-1}W_i\Delta_0\mathcal{J}^\mu\Delta_0W_{j-i}\right]
{\Pi}_0~.=\nonu =j^{c (n-1)\mu} +\overline\Pi_0
\left[
\sum_{i=0}^{n}W_i\Delta_0\mathcal{J}_0^\mu\Delta_0W_{n-i}
+\sum_{i=0}^{n-1}W_i\Delta_0\mathcal{J}_I^\mu\Delta_0W_{n-1-i} \right]
\Pi_0~,\label{truncurr}
\ee
where it has been formally defined $W_0\Delta_0=\Delta_0W_0= 1$. It is worth noting
that Eq. (\ref{truncurr}) contains  power of the interaction up to the $n^{th}$ order,
since   $\mathcal{J}_0^\mu$ is $O(V^0)$ and $\mathcal{J}_I^\mu$ is $O(V^1)$.

As demonstrated by induction in Appendix E, this truncated current
operator satisfies a WTI given by
\begin{eqnarray}
Q^\mu{j_\mu^{c (n)}}= {g_n}^{-1}~\widehat{\mathcal
Q}^L_{LF}-\widehat{\mathcal Q}^R_{LF}~{g_n}^{-1}~, \label{wtitrunc}
\end{eqnarray}
where  ${g_n}^{-1}={g_0}^{-1}-w^{(n)}$ and  $w^{(n)}$  is the truncated effective
interaction, viz
\begin{eqnarray}
w^{(n)}=  \overline{\Pi}_0 W^{(n)}{\Pi}_0=\sum_{i=1}^{n} ~
\overline{\Pi}_0 W_i {\Pi}_0.
  \label{wn}
\end{eqnarray}
 The matrix
elements of Eq.~(\ref{wtitrunc}) should be taken between solutions
of $g_n^{-1}|\phi_n\rangle=0$.

Thus, we conclude that the LF electromagnetic current operator
$j_\mu^{c (n)}$  is conserved at any given order $n$ of the QP
expansion. In the limit  $n\rightarrow \infty$ the truncated
conserved current becomes the full current operator of the model, if
the QP expansion converges (for numerical studies of the convergence
see e.g. \cite{sales00,miller3}). Finally, the effects  of the
truncation in the QP
expansion, and correspondingly in the Fock-space expansion,
on the transformation properties under the Poincar\'e group
will be investigated elsewhere \cite{MFPSSup}.

{ It should be pointed out that the last line of Eq. (\ref{truncurr}) suggests a possible
diagrammatic representation of the current (cf  the next Section
for the first-order case in a simple example), through the observation that the contribution to the
quasipotential at a $p^{th}$ order, $W_p$, can be decomposed
in $W_p=W_i~\Delta_0~W_{p-i}$ for $p\ge i\ge 0$ (see Eq. (\ref{Wn})). 
Indeed one can construct a picture of
the contribution between square brackets in the last line of Eq. 
(\ref{truncurr}), by starting with all the possible
decompositions for $W_n$ and $W_{n-1}$, then 
breaking the subtracted propagation represented by $\Delta_0$ and inserting the current
$\mathcal{J}_0$ and $\mathcal{J}_I$, respectively, 
i.e.  $\Delta_0 \to \Delta_0 \mathcal{J}_{0(I)}\Delta_0$.

 As shown in detail in the example of the next Section, the subtracted Green's function $\Delta_0$ 
eliminates the presence of
 disconnected  diagrams, due to the global, free-propagation  of
  the two-fermions
between the photon absorption and the exchange of a 
boson from the interaction. At the same time,
 it allows for the photon being absorbed i) instantaneously (in LF time!),
or ii)  by higher Fock-state 
components, including a
particle-antiparticle pair. 
In particular, one can easily see that  the disconnected diagrams  
generated from the gauging 
on the external legs, i.e. diagrams that are already taken into
account through the interacting LF wave function, are eliminated 
by the presence of the subtraction in
$\Delta_0$,  avoiding double counting.}

\section{LF Current and Ward-Takahashi Identity in the ladder Yukawa model
at the first order}

In this Section we explicitly derive  the current operator
$j^{c(1)\mu}$, see Eq.~(\ref{jcn1}), for a  two-fermion
system interacting through chargeless bosons, within the Yukawa model
in ladder approximation. This will
illustrate in an actual case how our procedure works to obtain
the conserved current operator at the first nontrivial order, namely
$n=1$.
The Yukawa model is defined by the following interacting Lagrangian
(with no derivative coupling):
\begin{eqnarray}
{\mathcal L}_I=g\overline\psi_1\Gamma_1^\alpha\psi _1\sigma_\alpha
+g\overline\psi_2\Gamma_2^\alpha\psi _2\sigma_\alpha , \label{li}
\end{eqnarray}
where $g$ is the
coupling constant (that should  not be confused with the LF Green's function
$g(k)$!), $\psi_1$ and $\psi_2$ are fermionic fields, the corresponding
particles have masses $m_1$ and $m_2 $, and charges $e_1$ and $e_2$,
respectively. The field $\sigma_\alpha$ corresponds to a chargeless
boson with mass $\mu$. The index $\alpha$ represents  Lorentz
components in the case of a vector or tensor field. The vertex
$\Gamma^\alpha_j$ is defined in the spinor space. { The four-dimensional interaction, $ V$
is given by \be  V=~ i~ g^2 ~{\Gamma_1^\alpha \otimes
\Gamma_{2\alpha} \over (\hat p_1 -\hat p_2)^2 -\mu^2+i\varepsilon}
\label{V4}\ee where the symbol $\otimes$ takes distinct the vertexes
corresponding to the fermions $1$ and $2$.}

\begin{figure}[thb!]
\centerline{\epsfig{figure=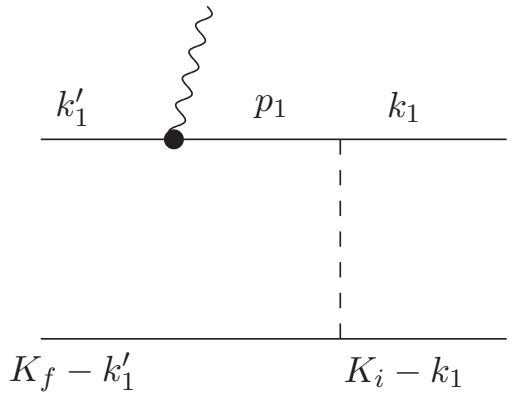,width=10cm}}
\caption{ Diagrammatic representation of the matrix elements of the
4-dimensional operator $ \overline G_0{\cal{J}}^\mu_0G_0V \overline G_0$, (cf. Eqs. (\ref{cvca2})
and (\ref{b2})). Dashed line:
four-dimensional interaction $V$. The full dot represents the free
current.}
\label{cladder0} 
\end{figure}

 We choose this simple example to show that an interacting LF
two-body current can be  generated even starting from a
four-dimensional free current, ${\cal{J}}^\mu_0$. It should be
pointed out that { in ladder BS approximation the consistent
four-current is the free one, i.e.  ${\cal{J}}^\mu \to
{\cal{J}}^\mu_0$, and  the four-dimensional WTI is fulfilled because
the commutator $[\widehat e,V]$
 trivially
vanishes. Such a cancellation is due to the fact that the exchanged
momentum in the potential just depends on the spectator particle
momentum, that remains unchanged in the transition from the initial
state to the final one (cf. Ref. \cite{gross} and Fig. \ref{cladder0}).}
Since we need all the four components of $j^{c(1)\mu}$  to construct
the WTI, Eq.~(\ref{wtijcn1}), in what follows we explicitly
evaluate all the components of the truncated em current,
by projecting the four-dimensional current onto the LF hypersurface
(i.e. by integrating on $k^-$).

Let us start with  the matrix elements of the
free current operator, given by
\be
\langle k_1|\mathcal{J}^{\mu}_0(Q)|p_1\rangle=~-2\pi
~e_1~\gamma^\mu_1~\delta^4(k_1-p_1-Q)~
\left [(\psla{K}_f-\psla{k}_1)-m_2)\right ]\nonu
   +\left [1 \rightarrow 2, k_1\rightarrow K_f-k_1,p_1\rightarrow
K_i-p_1\right ]~,\label{currfree0}
\end{eqnarray}
where $Q^\mu=K^\mu_f-K^\mu_i$. The factor  $(-2\pi)$ is introduced
in the current operator to make it compatible  with the free Green's
function, see Eq. (\ref{livre}).

The first-order current operator for the interacting two-fermion
system in this example is given by (cf Eq.~(\ref{jcn1})),
\begin{eqnarray}
{{j}^{c(1)\mu}}={{j}^{c(0)\mu}} +\overline
\Pi_0\left[V\Delta_0\mathcal{J}^\mu_0 +\mathcal{J}^\mu_0 \Delta_0V
\right]\Pi_0 \ , \label{cvca1}
\end{eqnarray}
where the zero order LF current operator is the LF free current, see
Eqs. (\ref{cvca0}) and (\ref{cvme2}).

The  contribution of the interaction to the LF current operator in
lowest order comes from two-body irreducible amplitudes given by the
second term in the r.h.s of Eq.~(\ref{jcn1}) with
$\mathcal{J}^\mu\equiv\mathcal{J}_0^\mu$, { since we are adopting
 the ladder approximation}. Using Eq.~(\ref{delta0}) for $\Delta_0$ we write
that:
\be
 {{j}^{c(1)\mu}}-{{j}^{c(0)\mu}}=\overline \Pi_0~V~G_0
 \mathcal{J}^\mu_0~\Pi_0
-w^{(1)}g_0{{j}^{c(0)\mu}}
  +\overline
\Pi_0~\mathcal{J}^\mu_0~G_0~V ~\Pi_0 -{{j}^{c(0)\mu}}g_0w^{(1)},
\label{cvca3} \ee {where $w^{(1)}=\overline \Pi_0 ~V ~\Pi_0$ is the
three-dimensional effective interaction, with matrix elements
 for a total momentum $K$ given by (see
Appendix C in \cite{sales01})
\be
\langle k_1^{\prime +}\vec{k}^\prime_{1\perp }|w^{(1)}(K)
|k_1^{+}\vec{k}_{1\perp }\rangle=~i~(i g)^2~\theta(k_1^{\prime +})\theta(k_1^{
+})\theta(K^+-k_1^{\prime +})\theta(K^+-k_1^{ +})\nonu \times
 \left[
\frac{\theta\left(k^{\prime+}_1-k^+_1\right)}{(k^{\prime+}_1-k^+_1)
\left ( K^- -k^{ -}_{1on}-k^{\prime-}_{2on}-k_{\sigma on} +i
\varepsilon\right )} + \frac{\theta\left(
k^+_1-k^{\prime+}_1\right)}{(k^+_1-k^{\prime+}_1) \left ( K^-
-k^{\prime -}_{1on}-k^{ -}_{2on}+k_{\sigma on} +i \varepsilon\right
)} \right] \nonu \times~
\Lambda_+(k^\prime_{1on})\Gamma^{\alpha}_1\Lambda_+(k_{1on})
\Lambda_+(k^\prime_{2on})\Gamma_{2\alpha}\Lambda_+(k_{2on}) \ .
\label{w1yukawa}\ee} where $k_2=K-k_1$, \be k^{ -}_{1on}={\vec k^{
2}_{1\perp}+m^2_{1}\over k^{ +}_{1}} \quad \quad \quad \quad
k^{\prime -}_{1on}={\vec k^{\prime 2}_{1\perp} +m^2_{1}\over
k^{\prime +}_{1}} \nonu k^{ -}_{\sigma on}={(\vec
k^\prime_{1\perp}-\vec k_{1\perp})^2+\mu^2 \over (k^{\prime +}_1-k^{
+}_1)}\ee { and the  quantities corresponding to fermion $2$ easily
follows.
 As  shown in Appendix F,  both the first term and the third
one in
the r.h.s. of Eq.~(\ref{cvca3}) contain a  LF two-body reducible part
(already taken into account when
${{j}^{c(0)\mu}}$ is applied to the  valence wave function), that
is canceled by the second and
the fourth ones, respectively.
In particular (see Ref. \cite{sales01}), the LF
two-body reducible contributions are generated by the
effective interaction
used to obtain the valence wave function
 at  the first order in the quasi-potential expansion of the
ladder BS equation. }   The essential role played by the reducible
terms was stressed in \cite{Ji}, in a calculation of higher Fock states
contributions to the Generalized Parton Distribution of pion.

The difference with the boson case, that was
recently studied \cite{adnei07}, comes from the instantaneous terms
present in the quasi-potential expansion of the fermionic current
operator,  as we will show in detail.

In the following we report the matrix elements of the first-order
current operator relevant for the em processes in the spacelike region,
as obtained in Appendix F.

\subsection{The term $\overline\Pi_0\mathcal{J}^\mu_0(Q)G_0V \Pi_0$ }

{ In Fig. 2, it is diagrammatically illustrated}
 the third term in the r.h.s. of Eq.~(\ref{cvca3}),
$g_0^{-1}|\overline G_0\mathcal{J}^\mu_0(Q)G_0V \overline
G_0|~g_0^{-1}$. { It is worthwhile to note that this term}
contains the relevant pair production contribution, for $Q^+>0$ (cf.
diagram (b) in Fig. 2). Differently, in the first term,
$g_0^{-1}|\overline G_0VG_0\mathcal{J}^\mu_0(Q) \overline
G_0|~g_0^{-1}$, there is no contribution from the pair production by
the virtual photon, since the conservation of the plus momentum
component always requires a positive
 plus component of
the intermediate fermion momentum and  then
the initial state cannot contain a photon and a pair
 {(cf. Fig. 3)}.

\begin{figure}[htb!]
\centerline{\epsfig{figure=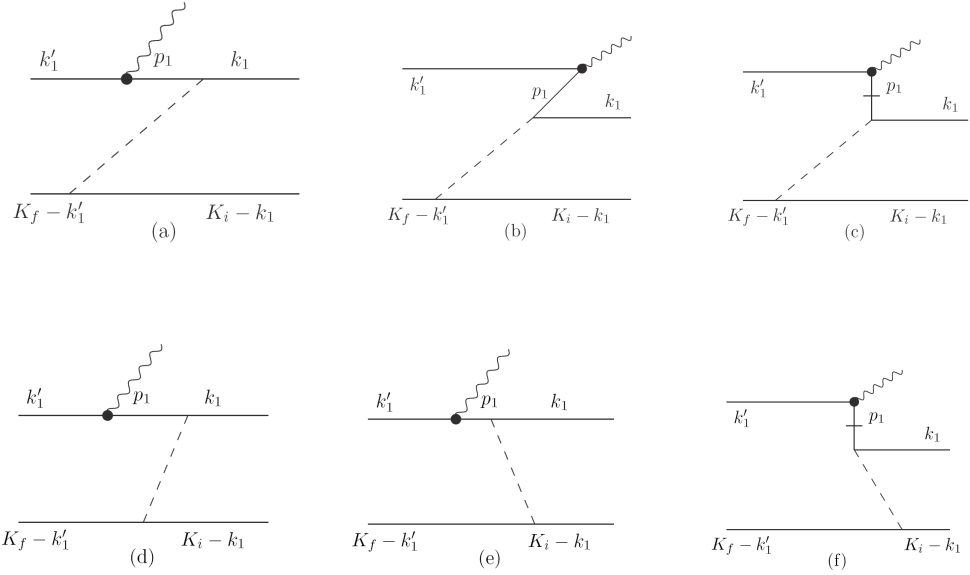,width=15cm}}
\caption{LF time-ordered diagrams representing  the matrix element
of the current operator obtained from $\overline
\Pi_0\mathcal{J}^\mu_0(Q)G_0V \Pi_0$, i.e. the third term
in the r.h.s. of Eq.~(\ref{cvca3}). The reducible diagrams, (d) and
(e), are canceled by ${{j}^{c(0)\mu}}g_0w^{(1)}$.}
\label{cladder1} 
\end{figure}

Performing analytical integrations over $k^-_1$ and $k^{\prime }_1$
 by Cauchy's theorem with the conditions $K_i^+ >0$
and $Q^+\geq 0$ (see Appendix F), we  get the six contributions
that appear in Fig. \ref{cladder1}. We observe that the limited
number of LF time ordered diagrams are essentially a consequence of
the trivial vacuum structure, due to the conservation of the total
positive plus momentum.  Now, let us
discuss in detail the irreducible diagrams of Fig.~\ref{cladder1}.

{ Diagram (a) gives a two-body current with  intermediate three-body
virtual state propagations before and after the photon absorption,
in the kinematical range $p^+_1=k^{\prime+}_1-Q^+\geq 0$ and
$k^+_1>(k^{\prime+}_1-Q^+)$. Diagram (b) represents the contribution
to the current of the virtual photon decay in a pair. The
kinematical condition in this case is $p^+_1=k^{\prime+}_1-Q^+<0$
(note that $-Q^+<p^+_1<0$).  Diagram (c) represents a contribution
from the instantaneous term of the fermion propagator, which is
non vanishing in  two kinematical regions:
 i) $p^+_1=k^{\prime+}_1-Q^+\geq 0$ and $k^+_1>(k^{\prime+}_1-Q^+)$, ii)
$p^+_1=k^{\prime+}_1-Q^+<0$.
Diagrams (d) and (e) are two-body reducible
terms which are canceled by ${{j}^{c(0)\mu}}g_0w^{(1)}$ in Eq. (\ref{cvca3}).
Diagram (f) represents a contribution of the
instantaneous term for $p^+_1=k^{\prime+}_1-Q^+ >k^+_1>0$.}

 Defining $\Theta=\theta(k^+_1)~\theta(k^{+\prime}_1)\theta(K^+_i -k^+_1)~\theta(K^+_f
-k^{+\prime}_1)$, the diagram (a) is given by (see Eq. (\ref{b5}))
\be
\langle k_1^{\prime +}\vec{k}_{1\perp }^{\prime }| j^{c(1)\mu}
|k_1^{+}\vec{k}_{1\perp }\rangle_{(a)}=  i e_1(2m_1)
(ig)^2~\Theta~
\frac{\theta\left(k^{+}_1-p^+_1\right)}{(k^+_1-p^+_1)p^+_1}
\frac{\theta\left(p^{+}_1\right)}{\Delta^{(1a)-}_f\Delta^{(1a)-}_i}
 \nonu
\times
\Lambda_+(k^\prime_{1on})\gamma^\mu_1\Lambda_+(p_{1on})\Gamma_1^\alpha
\Lambda_+(k_{1on})\Lambda_+((K_f-k^\prime_1)_{on})\Gamma_{2^\alpha}\Lambda_+((K_i-k_1)_{on})
 \ . \label{b5T}
\ee
The  four-vectors $\Delta^{(1a)}_f$ and $\Delta^{(1a)}_i$  are the following
combinations of the four-momenta
\be
 \Delta^{(1a)}_f=K_f-k^{\prime}_{1on} - (K_i-k_1)_{on}
-(k_1-p_1)_{on}+i\varepsilon \nonu \Delta^{(1a)}_i=K_i-
p_{1on}-(K_i-k_1)_{on} -(k_1-p_1)_{on} +i\varepsilon \ ,
\label{b6T}
\ee
The minus components  yield  the
three-body (2 fermions and the exchanged boson) global propagation in the final
and initial states, respectively.

The  matrix element of the current operator (b) (virtual photon decay in a pair)
is (see Eq. (\ref{b15}))
\be
 \langle k_1^{\prime +}\vec{k}_{1\perp }^{\prime }| j^{c(1)_\mu}
|k_1^{+}\vec{k}_{1\perp }\rangle_{(b)}=  -i e_1 (2m_1) (i g)^2
~\Theta~
{\theta\left(-p_1^+\right)\over p^+_1}
~\frac{\theta\left(k^{+}_1-p^+_1\right) }{(k^{+}_1-p^+_1 )
\Delta^{(1a)-}_f \Delta^{(1b)-}_\gamma} \nonu \times ~ \Lambda_+(k^\prime_{1on})
\gamma^\mu_1 \Lambda_+(p_{1on} )~\Gamma_1^{\alpha}\Lambda_+(k_{1on})
 \Lambda_+((K_f-k^{\prime}_1)_{on})\Gamma_{2\alpha}\Lambda_+((K_i-k_1)_{on})\ ,
 \label{b15T}
\ee
where the combination $\Delta^{(1b)}_\gamma$ of four momenta is
\be
\Delta^{(1b)}_\gamma=Q- k^\prime_{1on}+p_{1on} +i\varepsilon \ .
\label{b16t}
\ee

The instantaneous term (c) (see Eqs. (\ref{b5inst}) and (\ref{b15jc})
in Appendix F) can be expressed as
\be
\langle k_1^{\prime +}\vec{k}_{1\perp }^{\prime }| j^{c(1)\mu}
|k_1^{+}\vec{k}_{1\perp }\rangle_{(c)}=  \frac{i}{2} e_1
(ig)^2~\Theta~
{1\over p^+_1 }
\times\frac{\theta\left(k^{+}_1-p^+_1\right)}{(k^+_1-p^+_1)
\Delta^{(1a)-}_f}
\nonu \times ~
\Lambda_+(k^\prime_{1on})\gamma^\mu_1\gamma^+_1\Gamma^\alpha_1
\Lambda_+(k_{1on})\Lambda_+((K_f-k^\prime_1)_{on})\Gamma_{2\alpha}
\Lambda_+((K_i-k_1)_{on})
 \label{b5instT}
\ee

Finally the contribution from the
instantaneous term for $k^{\prime+}_1-Q^+>k^+_1$, represented by
diagram (f), is (see Eq. (\ref{b5f})):
\be
\langle k_1^{\prime +}\vec{k}_{1\perp }^{\prime }| j^{c(1)\mu}
|k_1^{+}\vec{k}_{1\perp }\rangle_{(f)}=  \frac{i}{2} e_1
(ig)^2~\Theta~
{\theta\left(p_1^+\right)\over p^+_1}~
\frac{\theta\left(p^{+}_1-k^+_1\right)}
{(p^+_1-k^+_1)
\Delta^{(1f)-}_i}\nonu \times ~
\Lambda_+(k^\prime_{1on})\gamma^\mu_1\gamma^+_1\Gamma_1^\alpha
\Lambda_+(k_{1on})\Lambda_+((K_f-k^\prime_1)_{on})\Gamma_{2^\alpha}\Lambda_+((K_i-k_1)_{on})
 \ , \label{b5fT}
\ee
where the denominator is the minus component of the following four-momentum
\be\Delta^{(1f)}_i=K_i-k_{1on}-(K_f-k^\prime_1)_{on}-(p_1-k_1)_{on}
\label{delta1fT}\ee
and yields  the intermediate propagation of the three-body system
composed by two fermions and a boson.

\begin{figure}[tb!]
\centerline{\epsfig{figure=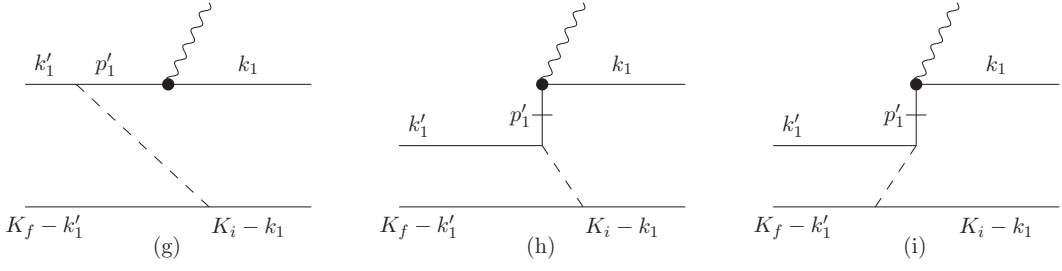,width=15cm}}
\caption{LF time-ordered diagrams representing  the matrix element
of the current operator obtained from $\overline
\Pi_0 V G_0\mathcal{J}^\mu_0(Q) \Pi_0$, i.e. the first term
in the r.h.s. of Eq.~(\ref{cvca3}). The reducible diagrams,
canceled by $w^{(1)} {j}^{c(0)\mu}g_0$, are not shown.}
\label{cladder2} 
\end{figure}

\subsection{The term $\overline\Pi_0VG_0 \mathcal{J}^\mu_0(Q)\Pi_0$}

Let us now discuss the first contribution in  Eq.~(\ref{cvca3}),
$g_0^{-1}|\overline G_0VG_0 \mathcal{J}^\mu_0(Q)\overline G_0|g_0^{-1}$,
depicted by
the diagrams (g) to (i) in Fig.~\ref{cladder2}.  It is worth noting
that the pair contribution is not present due to the conservation of
the plus momentum component. For the sake of simplicity, the two
reducible diagrams, which are canceled by
$w^{(1)}g_0{{j}^{c(0)\mu}}$ in Eq.~(\ref{cvca3}) in an analogous
way as we have already discussed for diagrams (e) and (d) in Fig.
\ref{cladder1}, are not
shown in Fig.~(\ref{cladder2}).
 As discussed in detail in Appendix F, by analytical integrations over $k^-_1$ and $k^{\prime }_1$
 by Cauchy's theorem one can express diagram (g)  as  (see Eq. (\ref{b8}))
\be
 \langle k_1^{\prime +}\vec{k}_{1\perp }^{\prime }| j^{c(1)\mu}
|k_1^{+}\vec{k}_{1\perp }\rangle_{(g)}=
  i e_1 (2m_1) (i g)^2~\Theta~
\frac{\theta\left(k^{\prime+}_1-p^{\prime+}_1\right)
}{(k^{\prime+}_1-p^{\prime+}_1 )p^{\prime+}_1
\Delta^{(1g)-}_f\Delta^{(1g)-}_i}
 \nonu
\times
\Lambda_+(k^\prime_{1on})\Gamma_1^\alpha\Lambda_+(p^\prime_{1on})\gamma^\mu_1
\Lambda_+(k_{1on})\Lambda_+((K_f-k^\prime_1)_{on})\Gamma_{2\alpha}\Lambda_+((K_i-k_1)_{on})
 \ ,
 \label{b8T}
\ee
where the combinations  $\Delta^{(1g)}_f$ and $\Delta^{(1g)}_i$ of
four momenta are
\begin{eqnarray}
&& \Delta^{(1g)}_f=K_f-p^\prime_{1on} - (K_f-k^\prime_1)_{on} -(
k^\prime_1-p^\prime_1)_{on}+i\varepsilon \nonumber \\ &&
\Delta^{(1g)}_i=K_i- k_{1on}-(K_f-k^\prime_1)_{on} -(
k^\prime_1-p^\prime_1 )_{on} +i\varepsilon = \Delta^{(1f)}_i\ .
\label{b9T}
\end{eqnarray}
{ since $k^\prime_1-p^\prime_1=p_1-k_1$.}

The  instantaneous diagram (h) is given by (see Eq. (\ref{b8inst}))
\begin{eqnarray}
 &&\langle k_1^{\prime +}\vec{k}_{1\perp }^{\prime }| j^{c(1)\mu}
|k_1^{+}\vec{k}_{1\perp }\rangle_{(h)}=
 \frac{i}{2} e_1  (i g)^2~\Theta~
\frac{\theta\left(k^{\prime+}_1-p^{\prime+}_1\right)
}{(k^{\prime+}_1-p^{\prime+}_1 )p^{\prime+}_1\Delta^{(1g)-}_i}
 \nonu
\times
\Lambda_+(k^\prime_{1on})\Gamma_1^\alpha\gamma^+_1\gamma^\mu_1
\Lambda_+(k_{1on})\Lambda_+((K_f-k^\prime_1)_{on})\Gamma_{2\alpha}
\Lambda_+((K_i-k_1)_{on})
 \ , \label{b8instT}
\end{eqnarray}
while the instantaneous contribution in the region $p_1^{\prime+}>k_1^{\prime+}> 0$,
illustrated by  diagram (i) in Fig. (\ref{cladder2}), is (see Eq. (\ref{b9inst}) )
\be
 \langle k_1^{\prime +}\vec{k}_{1\perp }^{\prime }| j^{c(1)\mu}
|k_1^{+}\vec{k}_{1\perp }\rangle_{(i)}=  \frac{i}{2} e_1  (i g)^2
~\Theta~\frac{\theta\left(p^{\prime+}_1-k^{\prime+}_1\right)
}{(p^{\prime+}_1-k^{\prime+}_1)p^{\prime+}_1\Delta^{(1i)-}_f}
 \nonu
\times
\Lambda_+(k^\prime_{1on})\Gamma_{1}^{\alpha}\gamma^+_1\gamma^\mu_1
\Lambda_+(k_{1on})\Lambda_+((K_f-k^\prime_1)_{on})\Gamma_{2\alpha}
\Lambda_+((K_i-k_1)_{on})
 \ , \label{b9instT}
\ee
where the three-body denominator is the minus component of
\begin{eqnarray}
\Delta^{(1i)}_f=K_f-k^\prime_{1on}-(K_i-k_1)_{on}-
(p^\prime_1-k^\prime_1)_{on}=\Delta_f^{(1a)} \ .
\label{delta1iT}\end{eqnarray}

\bigskip\bigskip

Closing the discussion on the first-order current, we refer the
reader to Appendix G for an explicit check of the current
conservation. The detailed derivation of current conservation allows
a deeper understanding of the explicit effects of the formal manipulations used to obtain
a conserved LF operator, consistent with the effective interaction
at each given order. This calculation  also shows
the essential role played by the instantaneous terms in the cancellation of terms from
 the two-body current (a) and (g),
that otherwise break current conservation. { In this respect it is
useful to consider that current (i), as current (c), yields
contribution in both the kinematical regions i) $p^+_1=k^{\prime
+}_1-Q^+>0$ and ii) $0>k^{\prime +}_1-Q^+=p^+_1$, as easily obtained
from the kinematical constraint $p_1^{\prime+}= k_1^+
+Q^+>k_1^{\prime+}$ leading to $ k_1^+> k_1^{\prime+}-Q^+=p^+_1 {>
\over <}0$.}

The peculiarity of the instantaneous terms in the treatment of the
fermionic case is also illustrated by the cancellation of logarithmic
singularities of the iterated ladder and stretched box diagrams of the
Yukawa model \cite{sales01,bakker07}. This cancellation yields the
full covariant form of the box diagram \cite{bakker07}.

\section{Conclusion}

In this paper we propose a conserved electromagnetic current
operator that acts on the valence component of the three-dimensional
LF wave function of a two-fermion system. In order to obtain the LF
current, {we have exploited  the quasi-potential approach to the BS equation
\cite{wolja} and then we have projected the relevant quantities onto the
light-front hyperplane.}  This approach has
been already applied to the BS equation for both boson and fermions
\cite{sales00,sales01} and to the construction of the conserved current
operator for two-boson systems~\cite{adnei07}.

The starting point of
the reduction scheme to the LF is  the QP approach to the BS equation with
 a proper  auxiliary four-dimensional Green's
function without instantaneous terms. Such terms, that represent the
peculiar feature of a fermionic system, are recovered both in the kernel
of the LF equation for the valence wave function and in the LF
two-body current operator.
The integration of the minus component of the particle four-momenta
in  the relevant operators, namely the Green's
function, the T-matrix and etc., allows one to accomplish the
desired LF projection of both eigenequation and em four-current. Furthermore,
we have identified a non-unitary operator, {\it the reverse LF-time projection operator},
 that acts on the LF
valence wave function and  allows one to reconstruct
 the four-dimensional BS amplitude
for bound or scattering states. Such an operator is  non unitary
since  the valence component does not carry the full normalization
of the wave function (for a discussion of the probability of the
higher Fock-components in the Wick-Cutkosky model see
Ref.~\cite{karmanovnpb}).  The explicit expression of the reverse
LF-time operator in terms of  the quasi-potential  is essential for
obtaining a LF three-dimensional current that fulfills the WTI, at
any order of the quasi-potential expansion.  In particular,  the
reverse LF time operator applied to a four-dimensional current
generates a three-dimensional LF current. Such an application
trivially leads to the equality between the matrix elements of the
four-current, evaluated by using the full BS amplitudes, and the
matrix elements, of the LF current evaluated by using the valence
wave functions corresponding to the previous BS amplitudes. Then the
current conservation  follows. This result is essential for
identifying the ingredients to be used for demonstrating the WTI for
the truncated LF current. In particular, we have defined the {\it
left} and {\it right} LF charge operators (that do not contain
interaction), and we have found the formal expression for the LF
WTI, where the LF Green's functions, for the initial and final
states, appear. The last step in our analysis, is of particular
relevance. Indeed,  a naive truncation of the QP expansion that
defines the LF current doesn't lead to a conserved current, while
retaining all the contributions up to a given order in the effective
interaction allows one to construct a LF truncated current that
satisfies the WTI. In such a truncated WTI, the truncated initial
and final Green's functions appear, as well as the same LF charge
operators obtained in the full case. In particular, the truncated
effective current operator can be put in correspondence to a sum
over intermediate states \cite{hierareq} up to some maximal number
of particles exchanged at a given LF time that flows from the
initial to the final valence states due to the photo-absorption
process. The truncation of the expansion of the quasi-potential
implies that the observables derived from the three-dimensional
conserved current operator are frame dependent,  but with  gauge
invariance correctly implemented.  The covariance under Lorentz
transformations of  the full LF current operator, $j^\mu(K_f,K_i)$,
is not greatly relevant, given the equality in Eq. (\ref{lfc1})
between the four-dimensional matrix elements of the covariant
current and the corresponding three-dimensional ones. However, since
the truncated theory is necessary for obtaining workable
approximations, the covariance properties of both the full and the
truncated LF current will be investigated elsewhere \cite{MFPSSup}.
Here, we can anticipate that  the covariance under the seven
kinematical LF transformations is satisfied by the LF current
operator (and also by the valence wave functions) in both  cases,
with a suitable introduction of new factors in the vertical bar
operation. For the truncated theory, the
 violation of covariance under dynamical LF transformations produces
 effects on the matrix elements of the current, that
can be reduced at any desired accuracy, by increasing the order $n$
of the truncated quasi-potential, namely approaching the full
theory. This  necessarily leads to consider  intermediate Fock
states with larger  and larger number of particles in the evaluation
of the effective interaction. Moreover, a quantitative study of the
effect of covariance violations was already performed in the
computation of the masses of bound states for the bosonic
model~\cite{sales00,miller3}. It was concluded that the expansion in
the Fock space is rapidly converging for a given covariant model
(see also \cite{miller1,miller2,carbonellepja}). In particular the
splitting between magnetic states of spin-1 composite bosons
decreases when the kernel of the LF bound state equation is
evaluated taking into account higher order terms, like stretched
boxes, in a bosonic model ~\cite{miller3}.

Our procedure for constructing a truncated LF current has been
illustrated in an actual case: the Yukawa model with chargeless boson exchange
in ladder
approximation. We have evaluated the LF current operator at the
lowest nontrivial order and have explicitly checked the
Ward-Takahashi identity for such model. The role of instantaneous terms
has been clarified and their relevance in producing  two-body
contributions has been emphasized.

{
Let us finally comment on possible problems about singularities and 
regularizations that can occur
in the truncated LF current.
Restricting to a model without self-energies and vertex corrections,  
but considering
in the kernel the ladder and two-body irreducible cross-ladder terms,
the Bethe-Salpeter equation is finite allowing a solution. 
However, the projection of this
equation to the light-front is plagued by infinities 
(see, e.g., \cite{miller1,miller2,Glazek,sales01,bakker07}).
 Recently
it was shown that the finite covariant box amplitude in the 
Yukawa model is fully recovered
when all terms in the light-front projection beyond the iterated box 
(i.e., stretched
 box and instantaneous terms) are obtained and that all the singularities are 
  canceled
 \cite{bakker07}. 
Thus in general one could argue that regularization problems occur in the
integration loops for the truncated current at order $n>1$, since the presence
in $\Delta_0$ of the global propagation, described by  $\widetilde G_0(K)$,
destroys the balance between singularities that allows for a finite result when
only $G_0(K)$ is considered (see e.g.\cite{bakker07}). For instance,
an easy form of regularization
can be introduced in $\widetilde G_0(K)$ through a cutoff function,
e.g. $\theta(\mu^2-M_0^2)$. Since the starting four-dimensional covariant 
BS model is finite, one expects that 
 the effect of the
scale $\mu$  vanishes if the QP expansion is not truncated.

However,  the issue of renormalization of the nonperturbative
bound state problem in the truncated QP expansion is subtle \cite{carbonell},
 and therefore the dependence upon  the scale $\mu$ in the three-dimensional 
 truncated
theory  should be carefully analyzed. This  non
trivial elaboration on the renormalization issue has to be postponed to a
future work.}

In summary, we have proposed a systematic expansion of a conserved
electromagnetic current operator within LF dynamics for two-fermion
interacting systems, using the quasi-potential approach to the
Bethe-Salpeter equation. As to the future perspectives, we plan to
apply such an interacting current for the investigation of inclusive
and exclusive electromagnetic processes, like hadron form factors and
deeply virtual photon scattering, {  after properly
generalizing the present approach to fermion-antifermion systems
 \cite{MFPSSup}.} It should be pointed out that
the identification of the matrix elements of any operator in the
four-dimensional space with the matrix elements of the corresponding
three-dimensional operator acting on the valence wave function is a
general procedure.

\section*{Acknowledgments}
This work was partially supported by the Brazilian agencies CNPq
and FAPESP and by Ministero della Ricerca Scientifica e
Tecnologica. J. A. O. M.  and T. F. acknowledge the hospitality of
the Dipartimento di Fisica, Universit\`a di Roma "Tor Vergata" and
of Istituto Nazionale di Fisica Nucleare, Sezione Tor Vergata and
 Sezione Roma I.


 \appendix

\section{Useful Identities}

We introduce the following identities that will be useful in
exploring the relation between LF and four-dimensional covariant quantities. It is
straightforward to get that:
\begin{eqnarray} (\psla p-m)(\psla
p_{on}+m)=(p^2-m^2)\frac{\gamma^+}{2p^+}(\psla p_{on}+m)~,
\label{pro1}
\\
(\psla p_{on}+m)(\psla p-m)=(p^2-m^2)(\psla
p_{on}+m)\frac{\gamma^+}{2p^+}~.\label{pro2}
\end{eqnarray}
These  identities imply  the following relations
\begin{eqnarray}
 \langle k^{\prime-} _1|G^{-1}_0(K)\overline G_0(K)| k^{-} _1
\rangle= 
\frac{\gamma_1^+}{2k_1^+}\frac{\gamma_2^+}{2k_2^+}(\psla
k_{1on}+m_1) (\psla k_{2on}+m_2)\delta(k^{\prime-} _1-k^{-} _1) \
, \label{I1}
\\
\langle k^{\prime-} _1|\overline G_0(K)G^{-1}_0(K)| k^{-} _1
\rangle= 
(\psla k_{1on}+m_1)  (\psla k_{2on}+m_2)\frac{\gamma_1^+}{2k_1^+}
\frac{\gamma_2^+}{2k_2^+} \delta(k^{\prime-} _1-k^{-} _1)\ .
\label{I2}
\end{eqnarray}
Note the presence of the product $\gamma_1^+\gamma_2^+$ that is essential for
the
definition of the LF charge operators, Eqs. (\ref{eilf}) and (\ref{eflf}).
Exploiting $(\gamma^+)^2=0$, one obtains another useful
identity
\begin{eqnarray}
(\psla p+m)\frac{\gamma^+}{2p^+}(\psla p+m)=(\psla
p_{on}+m),\label{pro3}
\end{eqnarray}
with $p^-_{on}=(\vec p^2_\perp +m)/p^+$.

Using i) Eqs.~(\ref{I1}) and  (\ref{pro3}), and ii) the explicit form of $\overline G_0(K)$,
one gets
\begin{eqnarray}
g_0(K)= |\overline G_0(K)G_0^{-1}(K)\overline G_0(K)| ~,
\label{Ag0}\end{eqnarray}
for the free three-dimensional propagator.

Now we can relate the interacting LF Green's function directly with
the four-dimensional Green's function by evaluating
\begin{eqnarray}
|\overline G_0(K)G_0^{-1}(K)G(K)G_0^{-1}(K)\overline G_0(K)|&=&
~|\overline G_0(K)G_0^{-1}(K)\overline G_0(K)|~+~|\overline
G_0(K)T(K)\overline G_0(K)|
\nonumber \\
&=&~ g_0(K)+g_0(K)\overline \Pi_0(K)T(K)\Pi_0(K)g_0(K)
\nonumber \\
&=&~g_0(K)+g_0(K)t(K)g_0(K)=g(K) ~,  \label{greenproj}
\end{eqnarray}
where Eq.~(\ref{3.1}) for the definition of $t(K)$ was used.

\section{The Interacting Reverse LF Projection Operator}
In this Appendix, we will prove some useful identities involving the interacting
reverse LF projection operator.

The following relation between $\Pi(K)$ (see the definition in Eq.
(\ref{invop})) and $T(K)$ can be
obtained from Eqs. (\ref{delta0}), (\ref{g0tilde}),
(\ref{LFRESOLV}),
(\ref{2.1}) and
  (\ref{3.1}). Indeed one has
 \be
 \Pi(K)={\Pi}_0(K)+\Delta_0(K)W(K){\Pi}_0(K)=\nonu=
 {\Pi}_0(K)+G_0(K)W(K){\Pi}_0(K) -{\Pi}_0(K)~g_0(K)\overline{\Pi}_0(K) W(K)
 {\Pi}_0(K)=
 \nonu=
 {\Pi}_0(K)+G_0(K)W(K){\Pi}_0(K) -{\Pi}_0(K)~g_0(K)w(K)
 \nonu={\Pi}_0(K)~g_0(K)g^{-1}(K)+G_0(K)W(K){\Pi}_0(K)~g(K)g^{-1}(K)=\nonu=
 \left[{\Pi}_0(K)+G_0(K)W(K){\Pi}_0(K)+G_0(K)W(K){\Pi}_0(K)
  g_0(K)t(K)\right]~g_0(K)~
 ~g^{-1}(K)=\nonu=\left[1+G_0(K)W(K)+G_0(K)W(K)\widetilde G_0(K)T(K)\right]
 {\Pi}_0(K)~g_0(K)~g^{-1}(K)=\nonu=
  \left[1+G_0(K)T(K)\right]{\Pi}_0(K)~g_0(K)~g^{-1}(K)
\label{rev1} \ee
Furthermore, by using Eqs. (\ref{gfull}) and (\ref{freeproj}), one gets
\be
\Pi(K)= G(K)~G^{-1}_0(K)~\overline G_0(K)|~g^{-1}(K)
 \ee
 that can be recast in the following form
 \be
G^{-1}(K)~\Pi(K)= G^{-1}_0(K)~\overline G_0(K)|~g^{-1}(K)
 \label{rev2}\ee
 The analogous expression for $\overline \Pi(K)$ reads
 \be
\overline \Pi(K)~ G^{-1}(K)= g^{-1}(K)~|\overline G_0(K) G^{-1}_0(K)
 \label{rev3}\ee

\section{zero-order LF Current operator and WTI}
{ In this Appendix we will
 evaluate  the matrix elements of the free LF current
operator, Eq.~(\ref{cvca0}), between free particle states and we will prove
explicitly the WTI.}
The matrix elements
of the
free 4-dimensional current operator are
\be
\langle k_1|\mathcal{J}^{\mu}_0(Q)|p_1\rangle=-2\pi~
\left[e_1~\gamma^\mu_1~\delta^4(k_1-p_1-Q)
~((\psla{K}_f-\psla{k}_1)-m_2)\right]\nonu
  +[1 \rightarrow 2, k_1\rightarrow K_f-k_1,p_1\rightarrow
K_i-p_1]~,
\ee
where $Q^\mu=K^\mu_f-K^\mu_i$. The factor  $(-2\pi)$ is introduced
in the current operator to make it compatible  with the free Green's
function, see Eq. (\ref{livre}).

 Since $g_0^{-1}$
is the identity in the two-particle space, modulo some factors, to
simplify the presentation let us consider
 in what follows only the relevant part of
${j^{c\mu}}^{(0)}$ containing the $k^-$-integration. Moreover,  in order
to make more fast
the discussion related to the position of the poles, we will take profit of the
$\theta$-functions present in the dropped  $g_0^{-1}$ factors, that
remind
us we
are dealing with particles in the external legs. Therefore, one has
 \be \Theta~
\langle k_1^{\prime +}\vec{k}_{1\perp }^{\prime }\left| \
~|\overline G_0 \mathcal{J}_0^\mu \overline G_0 |~ \ \right|
k_1^{+}\vec{k}_{1\perp }\rangle=  \nonu  - \frac{1 }{\left( 2\pi
\right) }~\Theta~ \int dk_1^{-}   \frac {e_1 \delta\left(k_1^{\prime
+}-k_1^+-Q^+\right)\delta^2\left(\vec k_{1\perp}^\prime-\vec k_{1
\perp}-\vec Q_{\perp}\right)}{k_1^{\prime+}k_1^{+}\left( k_1^{-}+Q^-
-k_{1on}^{\prime-} +i\frac{\varepsilon}{k_1^{\prime +}} \right)
\left( k_1^{-}-k_{1on}^{-} +i\frac{\varepsilon}{k_1^{+}}\right)}
\nonu \times   \frac {(\psla{k}^\prime_{1on}+m_1)\gamma_1^\mu
(\psla{k}_{1on}+m_1)((\psla{K}_i-\psla{k}_{1})_{on}+m_2)
}{(K_i^{+}-k_1^{+})\left( K_i^{-}-k_1^{-}-
(K_i-k_1)^-_{2on}+i\frac{\varepsilon}{K_i^{+}-k_1^{+}} \right) } \
+1\leftrightarrow 2, \label{cvme0} \ee where $k^\prime_1=k_1+Q$, and
  $\Theta=\theta(k^+_1)~\theta(k^{+\prime}_1)\theta(K^+_i -k^+_1)~\theta(K^+_f
-k^{+\prime}_1)$. It is important  noting that, without profiting of the presence of $\Theta$ on the left of
Eq. (\ref{cvme0}), a lengthy discussion of the
poles leads to the   presence of $\Theta$ in the
result.  We have used
the identities (\ref{pro1}) and (\ref{pro3}) to simplify the
following combination that appears in the numerator of
Eq.~(\ref{cvme0})
\begin{eqnarray}
(\psla{k}_{2on}+m_2) (\psla{k}_{2}-m_2)(\psla{k}_{2on}+m_2)=
(k^2_2-m^2_2)(\psla{k}_{2on}+m_2) \ ,
\end{eqnarray}
with $k_2=K_i-k_1=K_f-k^\prime_1$.

Integrating over $k^-_1$ and assuming that $K^+_i>0$ and $Q^+\geq
0$, without loss of generality, one  gets that
 \be \Theta~
\langle k_1^{\prime +}\vec{k}_{1\perp }^{\prime }\left| \
~|\overline G_0 \mathcal{J}_0^\mu \overline G_0 |~\ \right|
k_1^{+}\vec{k}_{1\perp }\rangle = i
\Theta ~ e_1\delta\left(k_1^{\prime +}-k_1^+-Q^+\right)\delta^2\left(\vec
k_{1\perp}^\prime-\vec k_{1 \perp}-\vec Q_{\perp}\right)\nonu
 \times  \frac
{(\psla{k}^\prime_{1on}+m_1)\gamma_1^\mu
(\psla{k}_{1on}+m_1)((\psla{K}_i-\psla{k}_{1})_{on}+m_2)
~}
{k_1^{\prime+} (K_i^{+}-k_1^{+})k_1^{+} \Delta^{(0)-}_f
\Delta^{(0)-}_i} \nonu
 +1\leftrightarrow 2
\ee
where we have introduced the four-vector quantities
$\Delta^{(0)}_f=K_f-k^\prime_{1on}-(K_f-k^\prime_1)_{on}+i\epsilon$
and $\Delta^{(0)}_i=K_i-k_{1on}-(K_i-k_1)_{on}+i\epsilon$ for
convenience. Note that the minus component of $\Delta_f$ and
$\Delta_i$ are the only non vanishing ones.
The on-minus-shell values of the individual momenta are
\begin{eqnarray}
&& k^{\prime-}_{1on}= \frac{\vec{k}^{\prime 2}
_{1\perp }+m_1^2}{k_1^{\prime+}}\\
&& k^-_{1on}= \frac{\vec{k}
_{1\perp }^2+m_1^2}{k_1^{+}}\\
&&(K_f-k^\prime_1)_{on}^-=\frac{(
\vec{K}_{f\perp }-\vec{k}^\prime_{1\perp
})^2+m_2^2}{K_f^{+}-k_1^{+}}
\\& &(K_i-k_1)_{on}^-=\frac{(
\vec{K}_{i\perp }-\vec{k}_{1\perp })^2+m_2^2}{K_i^{+}-k_1^{+}} \ .
\end{eqnarray}
Taking into account the definition of $g_0(K)$,
Eq.~(\ref{2.11a}), and the matrix elements of $\widehat e_{1LF}$,
Eq.~(\ref{d3}), we have
\be\langle k_1^{\prime +}\vec{k}_{1\perp }^{\prime }\left| \
~|\overline G_0 \mathcal{J}_0^\mu \overline G_0 |~\ \right|
k_1^{+}\vec{k}_{1\perp }\rangle=-i  \frac{K_i^{+}-k_1^{+}}{2m_2}\langle k_1^{\prime +},\vec
k^\prime_{1\perp}|g_0(K_f)\gamma_1^\mu~\widehat e_{1,LF} ~g_0(K_i)|
k_1^{+}\vec{k}_{1\perp}\rangle ~ +\nonu+1\leftrightarrow 2 \ ,
\label{cvme1}
\ee
Finally, by multiplying by the proper $g_0(K)^{-1}$, the
matrix element of the free current operator can be written as
\be
 \langle k_1^{\prime +}\vec{k}_{1\perp }^{\prime
}\left|{{j}^{c(0)\mu}} \right| k_1^{+}\vec{k}_{1\perp }\rangle
= -i\Theta
\Lambda_+(k^\prime_{1on})\gamma_1^\mu\Lambda_+(k_{1on})
\nonu \times \Lambda_+((K_i-k_1)_{on})
 \frac{K_i^{+}-k_1^{+}}{2m_2}\langle k_1^{\prime +},\vec
k^\prime_{1\perp}|\widehat e_{1,LF}|
k_1^{+}\vec{k}_{1\perp}\rangle
 +1\leftrightarrow 2  \label{cvme2b}
\ee

To derive the WTI for the free LF current operator, we evaluate the
four-divergence of the current by contracting it with $Q^\mu$. Since
$(K_f-k^\prime_1)_{on}^-=(K_i-k_1)_{on}^-$, which follows from
kinematical momentum conservation, the momentum transfer can be
written in terms of $\Delta$'s as,
$Q=\Delta^{(0)}_f-\Delta^{(0)}_i+k^\prime_{1on}-k_{1on}$, and thus
\begin{equation}
\Psla{Q}=\frac{\gamma^+}{2}
\left(\Delta^{(0)-}_f-\Delta^{(0)-}_i\right) \ . \label{slaq}
\end{equation}
Then one has
\begin{eqnarray}
\Lambda_+(k^\prime_{1on})\Psla{Q}\Lambda_+(k_{1on})=
\Lambda_+(k^\prime_{1on})\frac{\gamma_1^+}{2}\Lambda_+(k_{1on})
(\Delta^{(0)-}_f-\Delta^{(0)-}_i) \ .
\end{eqnarray}
>From the above results, the four-divergence of the free current
becomes:
\begin{eqnarray}
&& \langle k_1^{\prime +}\vec{k}_{1\perp }^{\prime }\left|Q\cdot
{{j}^{c}}^{(0)} \right| k_1^{+}\vec{k}_{1\perp }\rangle =
-i~\Theta~\Lambda_+(k^\prime_{1on})\frac{\gamma_1^+}{2}\Lambda_+(k_{1on})
\Lambda_+((K_i-k_1)_{on}) \nonumber \\ && \times
 \frac{K_i^{+}-k_1^{+}}{2m_2}(\Delta^{(0)-}_f-\Delta^{(0)-}_i)\langle k_1^{\prime +},\vec
k^\prime_{1\perp}|\widehat e_{1,LF}|
k_1^{+}\vec{k}_{1\perp}\rangle
 +1\leftrightarrow 2 ~. \nonumber \\
 && =\langle k_1^{\prime +}\vec{k}_{1\perp }^{\prime }\left|[g_0 (K_f)]^{-1}\widehat{\mathcal
Q}^L_{LF}-\widehat{\mathcal Q}^R_{LF}[g_0 (K_i)]^{-1}\right|
k_1^{+}\vec{k}_{1\perp }\rangle~,
\label{cvme3}
\end{eqnarray}
since the free resolvent that appears in Eq.~(\ref{cvme3}) comes
from the following relations
\begin{eqnarray}
&&\Delta^{(0)-}_i\Lambda_+(k_{1on})\Lambda_+(k_{2on})= i
\frac{2m_1}{k^+_1}\frac{2m_2}{K^+_i-k^+_1}\left[g_0(K_i)\right]^{-1}
\nonumber \\
&&\Delta^{(0)-}_f\Lambda_+(k_{1on}^\prime)\Lambda_+(k_{2on})= i
\frac{2m_1}{k^{\prime+}_1}\frac{2m_2}{K^+_f-k^{\prime+}_1}
\left[g_0(K_f)\right]^{-1} ~.
\end{eqnarray}
Equation (\ref{cvme3}) gives, as expected, the matrix elements of (\ref{divj0}), i.e.
the WTI for $n=0$.

\section{WTI for the  first-order LF current operator}

In this Appendix we show the WTI for LF current, obtained by truncating at
the first order the effective interaction, see Sect. VI. Let us rewrite the
LF first-order current operator
\begin{eqnarray}
j^{c(1)\mu}=\overline{\Pi}_0~\left[{\mathcal
J}^\mu_0+{\mathcal
J}^\mu_I+V\Delta_0{\mathcal J}^\mu_0+{\mathcal
J}_0^\mu\Delta_0V\right] ~ {\Pi}_0, \label{appb1}
\end{eqnarray}
 Using the  current conservation for $\mathcal{J}^\mu_0$ and
$\mathcal{J}^\mu_I$, see Eqs. (\ref{freewti}) and (\ref{jwti0}),
 the four-divergence  is given by
\be
Q^\mu{j_\mu^{c (1)}}={g_0}^{-1}\widehat{\mathcal
Q}^L_{LF}-\widehat{\mathcal Q}^R_{LF}{g_0}^{-1}+\overline{\Pi}_0
\left[(\hat{e}V-V\hat{e})
+V\Delta_0G_0^{-1}\hat{e} -\hat{e}G_0^{-1}\Delta_0V\right] {\Pi}_0~.
\ee
Note that the  term $[-V~\Delta_0~\hat{e}~G_0^{-1}~{\Pi}_0+
 \overline{\Pi}_0~G_0^{-1}
~\hat{e}~\Delta_0~V]$  is not present in the above equation, since it
is vanishing due to the absence of a Dirac structure in the operator
$\widehat e$
and because of Eqs. (\ref{I1}), (\ref{I2}) and (\ref{deltapp}).

{ Furthermore, by using the definitions of i) $\Delta_0$, Eq.
(\ref{delta0}), ii) $\widetilde G_0 $,   Eq. (\ref{g0tilde}), and
iii)
 $ \widehat{\mathcal Q}^L_{LF}$, Eq. (\ref{eilf}), and
 $ \widehat{\mathcal Q}^R_{LF}$, Eq.  (\ref{eflf}), one has
\be Q^\mu{j_\mu^{c (1)}}={g_0}^{-1}\widehat{\mathcal
Q}^L_{LF}-\widehat{\mathcal Q}^R_{LF}{g_0}^{-1}-\overline{\Pi}_0
\left[ V~\widetilde G_0~G_0^{-1}\widehat{e}
-\widehat{e}G_0^{-1}~\widetilde G_0~V\right] {\Pi}_0= \nonu=
{g_0}^{-1}\widehat{\mathcal Q}^L_{LF}-\widehat{\mathcal
Q}^R_{LF}{g_0}^{-1}-\overline{\Pi}_0 \left[ V~{\Pi}_0
~g_0~\overline{\Pi}_0~G_0^{-1}\hat{e} -\hat{e}G_0^{-1}~ {\Pi}_0~
g_0~\overline{\Pi}_0~V\right] {\Pi}_0=\nonu=
{g_0}^{-1}\widehat{\mathcal Q}^L_{LF}-\widehat{\mathcal
Q}^R_{LF}{g_0}^{-1}- \left[ w^{(1)}
~|\overline{G}_0~G_0^{-1}~\hat{e}~{\Pi}_0 -
\overline{\Pi}_0~\hat{e}~G_0^{-1}~ {G}_0|~w^{(1)}\right]
=\nonu=({g_0}^{-1}-w^{(1)})\widehat{\mathcal
Q}^L_{LF}-\widehat{\mathcal Q}^R_{LF}({g_0}^{-1}-w^{(1)})=
{g_1}^{-1}~\widehat{\mathcal Q}^L_{LF}-\widehat{\mathcal
Q}^R_{LF}~{g_1}^{-1}~. \ee}

\section{WTI for arbitrary $n>1$ LF current operator}

The proof of WTI is based on the induction hypothesis. Once the current
operator at order $n$ satisfies the Ward-Takahashi identity, we must
demonstrate its validity for $n+1$. First note that
\be
j^{c (n+1)\mu}={j^{c (n)\mu}} +
\overline\Pi_0\left[\vphantom{\sum_{i=1}^{n-1}}W_n\Delta_0\mathcal{J}^\mu_I+\mathcal{J}_I^\mu\Delta_0W_n+
W_{n+1}\Delta_0\mathcal{J}^\mu_0 +
\mathcal{J}_0^\mu\Delta_0W_{n+1}\right.\nonu\left. +
\sum_{i=1}^{n-1}W_i\Delta_0\mathcal{J}_I^\mu\Delta_0W_{n-i}
+\sum_{i=1}^{n}W_i\Delta_0\mathcal{J}_0^\mu\Delta_0W_{n-i+1} \right]
\Pi_0~,
\ee
where we have separated out the free term from the interacting one
in the 4-dimensional current using Eq.~(\ref{4dcurr}). Since
$Q_\mu\mathcal{J}_I^\mu=\hat{e}V-V\hat{e}$, by induction, we have:
\be
Q^\mu{j_\mu^{c (n+1)}}=~g_n^{-1}\widehat{\mathcal
Q}_{LF}^L-\widehat{\mathcal Q}_{LF}^R~g_n^{-1}+
\overline\Pi_0~\left[\vphantom{\sum_{i=1}^{n-1}}W_n\Delta_0(\hat{e}V-V\hat{e})+\right.
\nonu(\hat{e}V-V\hat{e})\Delta_0W_n+W_{n+1}\Delta_0(G_0^{-1}\hat{e}-
\hat{e}G_0^{-1})+(G_0^{-1}\hat{e}
-\hat{e}G_0^{-1})\Delta_0W_{n+1}+\nonu
\left.\sum_{i=1}^{n-1}W_i\Delta_0(\hat{e}V-V\hat{e})\Delta_0W_{n-i}
+\sum_{i=1}^{n}W_i\Delta_0(G_0^{-1}\hat{e}-\hat{e}G_0^{-1})\Delta_0W_{n-i+1}
\right]~\Pi_0~.
\ee
{ In the above equation, the term
$\overline\Pi_0~[-W_{n+1}\Delta_0\hat{e}G_0^{-1}+G_0^{-1}\hat{e}
\Delta_0W_{n+1}]~\Pi_0$
  vanishes due to the absence of a Dirac structure in the
operator $\widehat e$ and because of Eqs. (\ref{I1}), (\ref{I2}) and
(\ref{deltapp}). By using the same relations and the definition of
$\widetilde G_0$, Eq. (\ref{g0tilde}), we can simplify the following
term that appears in last sum, i.e. \be
\Delta_0(G_0^{-1}\hat{e}-\hat{e}G_0^{-1})\Delta_0
=(\hat{e}\Delta_0-\Delta_0\hat{e}) \ee} Then, since by definition
$V\Delta_0W_i=W_{i+1}$, we have
\be
Q^\mu{j_\mu^{c (n+1)}}=~g_n^{-1}\widehat{\mathcal
Q}_{LF}^L-\widehat{\mathcal Q}_{LF}^R~g_n^{-1}+
\overline{\Pi}_0\left[\vphantom{\sum_{i=1}^{n-1}}W_n\Delta_0\hat{e}V-V\hat{e}\Delta_0W_n
 -\right.\nonu W_{n+1}\tilde{G}_0G_0^{-1}\hat{e}
+\hat{e}G_0^{-1}\tilde{G}_0W_{n+1}+ 
\sum_{i=1}^{n-1}W_i\Delta_0\hat{e}W_{n-i+1}-
\sum_{i=1}^{n-1}W_{i+1}\hat{e}\Delta_0W_{n-i} +\nonu
\left.\sum_{i=1}^{n}W_i(\hat{e}\Delta_0-\Delta_0\hat{e})W_{n-i+1}
\right]~\Pi_0,
\ee
and rearranging the terms we get
\begin{eqnarray}
Q^\mu{j_\mu^{c (n+1)}}=g_n^{-1}~\widehat{\mathcal
Q}_{LF}^L-\widehat{\mathcal Q}_{LF}^R ~g_n^{-1}+
~\overline{\Pi}_0\left[ \vphantom{\sum_{i=1}^{n-1}}\hat{e}G_0^{-1}\tilde{G}_0W_{n+1}
-W_{n+1}\tilde{G}_0G_0^{-1}\hat{e} +\right.\nonumber\\ \left.
\sum_{i=1}^{n}W_i\Delta_0\hat{e}W_{n-i+1}-
\sum_{i=0}^{n-1}W_{i+1}\hat{e}\Delta_0W_{n-i}+
\sum_{i=1}^{n}W_i(\hat{e}\Delta_0-\Delta_0\hat{e})W_{n-i+1}\right]
\Pi_0~
\end{eqnarray}
Since the sums cancel each other, we have
\be
Q^\mu{j_\mu^{c (n+1)}}=g_n^{-1}~\widehat{\mathcal
Q}_{LF}^L-\widehat{\mathcal Q}_{LF}^R ~g_n^{-1}+
~\overline{\Pi}_0\left[ \hat{e}G_0^{-1}\tilde{G}_0W_{n+1}
-W_{n+1}\tilde{G}_0G_0^{-1}\hat{e}\right]\Pi_0=\nonu =
g_n^{-1}~\widehat{\mathcal
Q}_{LF}^L-\widehat{\mathcal Q}_{LF}^R ~g_n^{-1}+
\widehat{\mathcal
Q}_{LF}^R w_{n+1} -w_{n+1}\widehat{\mathcal
Q}_{LF}^L \ee
where the LF
charge operators as given by Eqs. (\ref{eflf}) and (\ref{eilf}) have been
introduced through
\be
 \overline{\Pi}_0
\hat{e}G_0^{-1}\tilde{G}_0W_{n+1}~\Pi_0=\widehat{\mathcal
Q}_{LF}^R w_{n+1}  \nonu   \overline{\Pi}_0
W_{n+1}\tilde{G}_0G_0^{-1}\hat{e}\Pi_0=w_{n+1}\widehat{\mathcal
Q}_{LF}^L  \ ,
\ee
Finally, since $ g_{n+1}^{-1}=g_{n}^{-1}-w_{n+1}$, one gets
\begin{eqnarray}
Q^\mu{j_\mu^{c (n+1)}}=g_{n+1}^{-1}\widehat{\mathcal
Q}_{LF}^L-\widehat{\mathcal Q}_{LF}^R~g_{n+1}^{-1}~.
~\label{truncwti}
\end{eqnarray}
Thus, by induction, we conclude that the LF electromagnetic current
operator $j_\mu^{c (n)}$  is conserved at any given order $n$ of the
quasi-potential expansion, once the matrix elements are taken
between eigenstates of: $g_{n}^{-1}|\phi_n\rangle=0$.

\section{Interaction-dependent part of the LF first-order current operator in ladder approximation}

The  contribution of the interaction to the LF current operator in
lowest order in ladder approximation comes from two-body irreducible amplitudes given by the
second term in the r.h.s of Eq.~(\ref{cvca1}).
 Using Eq.~(\ref{delta0}) for $\Delta_0$ one has:
\be
 {{j}^{c(1)\mu}}-{{j}^{c(0)\mu}}=\overline \Pi_0~V~G_0
 \mathcal{J}^\mu_0~\Pi_0
-w^{(1)}g_0{{j}^{c(0)\mu}}
  +\overline
\Pi_0~\mathcal{J}^\mu_0~G_0~V ~\Pi_0 -{{j}^{c(0)\mu}}g_0w^{(1)},
\label{cvca2} \ee { where $w^{(1)}=\overline \Pi_0 ~V ~\Pi_0$ is the
three-dimensional effective interaction (see Eq.~(\ref{w1yukawa})).}

In the following we analyze the matrix elements of the first-order
current operator relevant for the em processes in the spacelike region.

\subsection{Evaluation of the term $|\overline
G_0\mathcal{J}^\mu_0(Q)G_0V \overline G_0|$ }

{ Let us start with}
 the third term in the r.h.s. of Eq.~(\ref{cvca2}),
$g_0^{-1}|\overline G_0\mathcal{J}^\mu_0(Q)G_0V \overline
G_0|~g_0^{-1}$, { which is diagrammatically illustrated in Fig. 2}.
{ It is important to observe that in this term}
contains the relevant pair production contribution, for $Q^+>0$ (cf.
diagram (b) in Fig. 2).
 { For the sake
of simplicity, we can first drop the multiplicative factors $g_0^{-1}$ on the
left and on the right, and we will consider  the current only for
particle 1. However, we can take advantage of the $\theta$-functions contained
in both $g_0^{-1}(K_f)$ and $g_0^{-1}(K_i)$, that greatly help in the
discussion of the analytical integrations over $k^{ -}_1$ and $k^{\prime -}_1$.}
 Therefore, we have to evaluate the following
matrix element { between free particle states} \be \Theta~ \langle
k_1^{\prime +}\vec{k}_{1\perp }^{\prime }|~\left| \overline G_0
\mathcal{J}^\mu_0(1) G_0 V\overline G_0\right|~
|k_1^{+}\vec{k}_{1\perp }\rangle= i e_1 \left(\frac{i g }{
2\pi}\right)^2~\Theta \nonu \times\int dk_1^{\prime -}dk_1^{-}
 \frac 1{k_1^{\prime +}(K_f^{+}-k_1^{\prime +})}\frac
{\psla{k}^\prime_{1on}+m_1}{\left(
k_1^{\prime -}- k^{\prime-}_{1on} +i{\varepsilon \over k_1^{\prime +}}
\right) }\frac{(\psla{K}_f-\psla{k}^\prime_1)_{on}+m_2}
{\left( K_f^{-}-k_1^{\prime -}-(K_f-k_1^{\prime})^-_{on}
+i{\varepsilon \over K_f^{+}-k_1^{\prime
+}}\right) } \nonu
\times ~ \gamma^\mu_1 \frac {\psla{p}_{1}+m_1}{p_1^{+}\left(
 p_1^{-}- p^{-}_{1on}  +i{\varepsilon \over p_1^{+}}\right)}
\frac { \Gamma_1^\alpha\Gamma_{2\alpha}}{(k_1^{+}-p_1^{+} )\left(
k_1^{-}-p_1^{-}-(k_1-p_1)^-_{on}+i{\varepsilon \over
k_1^{+}-p_1^{+}}\right) } \nonu \times ~\frac {
1}{k_1^{+}(K_i^{+}-k_1^{+})} \frac {\psla{k}_{1on}+m_1}{\left(
k_1^{-}-k^{-}_{1on} +i {\varepsilon \over k_1^{+}}\right) } \frac
{(\psla{K}_i-\psla{k}_1)_{on}+m_2}{ \left( K_i^{-}-k_1^{-}-
(K_i-k_1)^-_{on} +i{\varepsilon \over K_i^{+}-k_1^{+}} \right) }\ ,
\label{b2} \ee where
$\Theta=\theta(k^+_1)~\theta(k^{+\prime}_1)\theta(K^+_i
-k^+_1)~\theta(K^+_f -k^{+\prime}_1)$ and
$p^\mu_1=k_1^{\prime\mu}-Q^\mu$. Eq.~(\ref{b2}) is represented by
the Feynman diagram shown in Fig.~\ref{cladder0}. { The six poles in
Eq.~(\ref{b2}) are \be k_{1A}^{-}=k^{-}_{1on} -i {\varepsilon \over
k_1^{+}}\nonu k_{1B}^{-}=K_i^{-}- (K_i-k_1)^-_{on} +i{\varepsilon
\over K_i^{+}-k_1^{+}} \nonu
 k_{1C}^{-}=p_1^{-}+(k_1-p_1)^-_{on}-i{\varepsilon \over k_1^{+}-p_1^{+}}\nonu
k_{1A}^{\prime -}= k^{\prime-}_{1on} -i{\varepsilon \over k_1^{\prime +}}\nonu
k_{1B}^{\prime -}=K_f^{-}-(K_f-k_1^{\prime-})^-_{on}
+i{\varepsilon \over K_f^{+}-k_1^{\prime +}}\nonu
p_1^{-}=k_{1C}^{\prime -}-Q^-= p^{-}_{1on}  -i{\varepsilon \over p_1^{+}}
\label{pole3}\ee
 with
  the on-minus-shell definition of the respective momenta given by
\be
k^{-}_{1on}=\frac{\vec{k}_{1\perp }^{
2}+m_1^2}{k_1^{+}} \nonu
(K_i-k_1)^-_{on}=\frac{(\vec{K}_{i\perp }-\vec{k}_{1\perp
})^2+m_2^2}{K_i^{+}-k_1^{+}} \nonu
(k_1-p_1)^-_{on}=\frac{( \vec{k}_{1\bot}-\vec{p}_{1\bot
}) ^2+\mu ^2}{k_1^{+}-p_1^{+}}
\nonu p^-_{1on}=\frac{\vec{p}_{1\perp }^{2}+m_1^2}{p_1^{+}}
\nonu k^{\prime-}_{1on}=\frac{\vec{k}_{1\perp }^{\prime
2}+m_1^2}{k_1^{\prime +}} \nonu
(K_f-k_1^{\prime})^-_{on}=\frac{(\vec{K}_{f\perp }-\vec{k}^{\prime}_{1\perp
})^2+m_2^2}{K_f^{+}-k_1^{\prime+}}\ .
\label{b7}
\ee}

The integrations over $k^-_1$ and $k^{\prime }_1$ in Eq.~(\ref{b2}) are performed
analytically using Cauchy's theorem with the conditions $K_i^+ >0$
and $Q^+\geq 0$. As a result, we  get the six contributions
that appear in Fig. \ref{cladder1}.
 Note that diagrams (d) and (e) are two-body reducible
terms which are canceled by ${{j}^{c(0)\mu}}g_0w^{(1)}$ in Eq. (\ref{cvca2}). Let us now
discuss in detail the diagrams of Fig.~\ref{cladder1}, devoting a specific
subsection to the pair diagram (b).

\subsubsection{ Diagrams (a), (c), (d), (e) and (f) for $p^+_1=k^{\prime+}_1-Q^+\geq
0$}
 { Let us start considering the kinematical region
$p^+_1=k^{\prime+}_1-Q^+\geq 0$ and $k^+_1>k^{\prime+}_1-Q^+$. We first
perform  the analytical integration of Eq.~(\ref{b2}) on $k^-_1$. The poles
$k^-_{1A}$ and  $k^-_{1C}$ belong to the lower semi-plane,
while $k^-_{1B}$ lays in the upper semi-plane.
  The result obtained integrating in the upper semi-plane
  contains contributions  corresponding to  diagrams (a), (c)  and (d), viz
\be
\Theta~\langle k_1^{\prime +}\vec{k}_{1\perp }^{\prime }|~\left|
\overline G_0 \mathcal{J}^\mu_0(1) G_0 V\overline G_0\right|~
|k_1^{+}\vec{k}_{1\perp }\rangle_{(a)+(c)+(d)}=  e_1
\frac{(i g)^2}{ 2\pi}~~
\Theta~
\theta\left(p^{+}_1\right)
\theta\left(k^{+}_1-p^+_1\right)
 \nonu
\times \int dk_1^{\prime -} \frac 1{k_1^{\prime
+}(K_f^{+}-k_1^{\prime +})}\frac{\psla{k}^\prime_{1on}+m_1}{\left(
k_1^{\prime -}-k^{\prime -}_{1on} +i{\varepsilon \over k_1^{\prime +}}
\right) }\frac {(\psla{K}_f-\psla{k}^\prime_1)_{on}+m_2}
{\left[ K_f^{-}-k_1^{\prime -}- (K_f-k_1^{\prime})^-_{on} +i{\varepsilon
\over K_f^{+}-k_1^{\prime
+}}\right] } \nonu \times \frac 1{(k_1^{+}-p_1^{+}
)p_1^{+}}\frac {\gamma^\mu_1}{\left[
K_f^{-}-k_1^{\prime-}- (K_i-k_1)^-_{on}+i{\varepsilon \over K_i^{+}-k_1^{+}}
- (k_1-p_1)^-_{on} +i{\varepsilon \over k_1^{+}-p_1^{+}}\right] } \nonu\times
\left[\frac {\psla{p}_{1on}+m_1}{\left(
 k_1^{\prime-}-Q^- - p^-_{1on}+i{\varepsilon \over p_1^{+}}\right) }
 +\frac{\gamma^+_1}{2}\right]
\frac
{\Gamma_1^\alpha\Gamma_{2\alpha}}{k_1^{+}(K_i^{+}-k_1^{+})}\nonu\times\frac
{(\psla{k}_{1on}+m_1)~[(\psla{K}_i-\psla{k}_1)_{on}+m_2]}{\left[
K_i^{-}- (K_i -k_1)^-_{on}+i{\varepsilon \over K_i^{+}-k_1^{+}}
-k_{1on}^{-} +{i\varepsilon \over k_1^{+}}\right] }  \ .
\label{b3}
\ee}
Note that i) the instantaneous term, proportional to
$\gamma^+_1$,  leading to a first contribution to the  current (c) shown in Fig. \ref{cladder1},
is explicitly separated out in Eq. (\ref{b3}),
ii) the global propagation
 of the initial state, i.e. $1/\left [K_i^{-}- (K_i -k_1)^-_{on}
-k_{1on}^{-}\right]$ will be canceled by $g_0(K_i)^{-1}$.

In order to separate the processes corresponding to diagrams (a) and
(d) of Fig. \ref{cladder1}, we make use of the identity
\be
 \frac 1 {\left(
K_f^{-}-k_1^{\prime-}- (K_i-k_1)^-_{on}+i{\varepsilon \over K_i^{+}-k_1^{+}}
- (k_1-p_1)^-_{on} +i{\varepsilon \over k_1^{+}-p_1^{+}}\right) }
 \frac 1{\left(
 k_1^{\prime-}-Q^-- p^-_{1on}+i{\varepsilon \over p_1^{+}} \right)}
 =   \nonu
  \left[\frac 1 {\left(
K_f^{-}-k_1^{\prime-}- (K_i-k_1)^-_{on}+i{\varepsilon \over K_i^{+}-k_1^{+}}
- (k_1-p_1)^-_{on} +i{\varepsilon \over k_1^{+}-p_1^{+}}\right) }
+ \frac 1{\left(
 k_1^{\prime-}-Q^-- p^-_{1on}+i{\varepsilon \over p_1^{+}} \right)}
\right] \nonu \times \frac1{K^-_i- p^-_{1on}+i{\varepsilon \over
p_1^{+}}- (K_i-k_1)^-_{on}+ i{\varepsilon \over K_i^{+}-k_1^{+}} -
(k_1-p_1)^-_{on} +i{\varepsilon \over k_1^{+}-p_1^{+}}} \ ,
\label{fracpar} \ee and integrate Eq.~(\ref{b3}) analytically on {
$k^{\prime-}_1$}. {
 The first term in the square brackets
in Eq.~(\ref{fracpar}) generates the contribution illustrated by diagram (a) in Fig.
(\ref{cladder1}), once we take the residue at the pole $k^{\prime}_{1A}$, in the
lower semi-plane, and
multiply   on the left by $g_0(K_f)^{-1}$ and on the right by
$g_0(K_i)^{-1}$ (cf Eq. (\ref{2.11b})).  The second  term in
the square brackets leads to the contribution of diagram (d),
  a two-body reducible term,  which is  canceled out
by one of the contributions in ${{j}^{c(0)\mu}}g_0w^{(1)}$, see  Eq.~(\ref{cvca2}).
 The diagram (a) is given by
\be
\langle k_1^{\prime +}\vec{k}_{1\perp }^{\prime }| j^{c(1)\mu}
|k_1^{+}\vec{k}_{1\perp }\rangle_{(a)}=  i e_1(2m_1)
(ig)^2~\Theta~
\frac{\theta\left(k^{+}_1-p^+_1\right)}{(k^+_1-p^+_1)p^+_1}
\frac{\theta\left(p^{+}_1\right)}{\Delta^{(1a)-}_f\Delta^{(1a)-}_i}
 \nonu
\times
\Lambda_+(k^\prime_{1on})\gamma^\mu_1\Lambda_+(p_{1on})\Gamma_1^\alpha
\Lambda_+(k_{1on})\Lambda_+((K_f-k^\prime_1)_{on})\Gamma_{2^\alpha}\Lambda_+((K_i-k_1)_{on})
 \ . \label{b5}
\ee
The  four-vectors $\Delta^{(1a)}_f$ and $\Delta^{(1a)}_i$  are the following
combinations of the four-momenta
\be
 \Delta^{(1a)}_f=K_f-k^{\prime}_{1on} - (K_i-k_1)_{on}
-(k_1-p_1)_{on}+i\varepsilon \nonu \Delta^{(1a)}_i=K_i-
p_{1on}-(K_i-k_1)_{on} -(k_1-p_1)_{on} +i\varepsilon \ ,
\label{b6}
\ee
The minus components  yield  the
three-body (2 fermions and the exchanged boson) global propagation in the final
and initial states, respectively.}

The instantaneous term  (c), in the kinematical region
under consideration in this subsection, can be obtained from the same pole $k^{\prime}_{1A}$.
Then one has
\be
\langle k_1^{\prime +}\vec{k}_{1\perp }^{\prime }| j^{c(1)\mu}
|k_1^{+}\vec{k}_{1\perp }\rangle^I_{(c)}=  \frac{i}{2} e_1
(ig)^2~\Theta~
{\theta\left(p_1^+\right)\over p^+_1 }
\times\frac{\theta\left(k^{+}_1-p^+_1\right)}{(k^+_1-p^+_1)
\Delta^{(1a)-}_f}
\nonu \times ~
\Lambda_+(k^\prime_{1on})\gamma^\mu_1\gamma^+_1\Gamma^\alpha_1
\Lambda_+(k_{1on})\Lambda_+((K_f-k^\prime_1)_{on})\Gamma_{2\alpha}
\Lambda_+((K_i-k_1)_{on})
 \label{b5inst}
\ee

Let us then consider the kinematical region where
$p^+_1=k^{\prime+}_1-Q^+ >k^+_1$. {  We first perform the analytical
integration of Eq.~(\ref{b2}) in $k^-_1$ by calculating the residue
at the pole $k^-_{1A}$
 in the lower semi-plane, where only this pole appears.
 Then in
this region we obtain two contributions from Eq. (\ref{b2}): a
reducible one (cf diagram (e) in Fig.~(\ref{cladder1})) and an
irreducible one given by diagram (f), namely an instantaneous term.
Diagram (e) is canceled by the remaining part of
${{j}^{c(0)\mu}}g_0w^{(1)}$.
The result corresponding to diagram (f) is}
\be
\Theta~\langle k_1^{\prime +}\vec{k}_{1\perp }^{\prime }|\left|
\overline G_0 \mathcal{J}^\mu_0(Q) G_0 V\overline G_0\right|
|k_1^{+}\vec{k}_{1\perp }\rangle_{(f)}=   e_1
\frac{(i g)^2}{ 2\pi}~~
\Theta~
\theta\left(p^{+}_1\right) \theta\left(
p^+_1-k^{+}_1\right)
 \nonu
\times \int dk_1^{\prime -} \frac 1{k_1^{\prime
+}(K_f^{+}-k_1^{\prime +})}\frac{\psla{k}^\prime_{1on}+m_1}{\left(
k_1^{\prime -}- k_{1on}^{\prime -}  +i{\varepsilon \over k_1^{\prime +}}
\right) }\frac {(\psla{K}_f-\psla{k}^\prime_1)_{on}+m_2}
{\left[ K_f^{-}-k_1^{\prime -}-( K_f-k_1^{\prime })^-_{on}+
i{\varepsilon \over K_f^{+}-k_1^{\prime
+}}\right] } \nonu \times \frac 1{(p_1^{+}-k_1^{+}
)p_1^{+}}\frac {\gamma^\mu_1}{\left[
k_1^{\prime-}-Q^- - k_{1on}^{-} -(p_1-k_1)_{on}^{-}+i{\varepsilon \over k_1^{+}}
+i{\varepsilon\over
p_1^{+}-k_1^{+}}\right] } \nonu\times
\frac{\gamma^+_1}{2} \frac
{\Gamma_1^\alpha\Gamma_{2\alpha}}{k_1^{+}(K_i^{+}-k_1^{+})}\frac
{(\psla{k}_{1on}+m_1)((\psla{K}_i-\psla{k}_1)_{on}+m_2)}{\left[
K_i^{-}-(K_i-k_1)_{on}^{-} +i{\varepsilon\over K_i^{+}-k_1^{+}}-k_{1on}^-
+i{\varepsilon\over k_1^{+}}\right] }  \ .
\label{b3f}
\ee

The analytical integration of Eq.~(\ref{b3f}) is easily performed  by
calculating the residue at the pole
$ k_{1B}^{\prime -}$ in the upper semi-plane.  Therefore, the
contribution from the
instantaneous term for $k^{\prime+}_1-Q^+>k^+_1$, represented by
diagram (f), is:
\be
\langle k_1^{\prime +}\vec{k}_{1\perp }^{\prime }| j^{c(1)\mu}
|k_1^{+}\vec{k}_{1\perp }\rangle_{(f)}=  \frac{i}{2} e_1
(ig)^2~\Theta~
{\theta\left(p_1^+\right)\over p^+_1}~
\frac{\theta\left(p^{+}_1-k^+_1\right)}
{(p^+_1-k^+_1)
\Delta^{(1f)-}_i}\nonu \times ~
\Lambda_+(k^\prime_{1on})\gamma^\mu_1\gamma^+_1\Gamma_1^\alpha
\Lambda_+(k_{1on})\Lambda_+((K_f-k^\prime_1)_{on})\Gamma_{2^\alpha}\Lambda_+((K_i-k_1)_{on})
 \ , \label{b5f}
\ee
where in the denominator appears the minus component of the following four-momentum
\be\Delta^{(1f)}_i=K_i-k_{1on}-(K_f-k^\prime_1)_{on}-(p_1-k_1)_{on}
\label{delta1f}\ee
which yields  the intermediate propagation of the three-body system
composed by two fermions and a boson.

\subsubsection{ Diagrams (b) and (c) for $0\geq k^{\prime +}_1- Q^+=p_1^+$}
{ In the kinematical region where $0\geq k^{\prime +}_1- Q^+=p_1^+$
 the non vanishing result can be
 obtained from the pole $k^-_{1B} $ in the upper semi-plane, as in Eq.
(\ref{b3}).  Then one has to consider the pole
$k^{\prime -}_{1A}$, when the integration over $k^{\prime -}_1$ is performed.}
 Note that in Eq.~(\ref{b3}) $\theta(p^+_1)$
has  now to be substituted by $\theta(-p^+_1)$.
The resulting matrix element of the current operator (b) is
\be
 \langle k_1^{\prime +}\vec{k}_{1\perp }^{\prime }| j^{c(1)\mu}
|k_1^{+}\vec{k}_{1\perp }\rangle_{(b)}= - i e_1 (2m_1) (i g)^2
~\Theta~
{\theta\left(-p_1^+\right)\over p^+_1}
~\frac{\theta\left(k^{+}_1-p^+_1\right) }{(k^{+}_1-p^+_1 )
\Delta^{(1a)-}_f \Delta^{(1b)-}_\gamma} \nonu \times ~ \Lambda_+(k^\prime_{1on})
\gamma^\mu_1 \Lambda_+(p_{1on} )~\Gamma_1^{\alpha}\Lambda_+(k_{1on})
 \Lambda_+((K_f-k^{\prime}_1)_{on})\Gamma_{2\alpha}\Lambda_+((K_i-k_1)_{on})\ , \label{b15}
\end{eqnarray}
where the combination $\Delta^{(1b)}_\gamma$ of four momenta is
\begin{eqnarray}
\Delta^{(1b)}_\gamma=Q- k^\prime_{1on}+p_{1on} +i\varepsilon \ .
\label{b16}
\ee

Finally one has to compute the two-body current due to the
instantaneous term (c), which in this kinematical region is given by:
\be
 \langle k_1^{\prime +}\vec{k}_{1\perp }^{\prime }| j^{c(1)_\mu}
|k_1^{+}\vec{k}_{1\perp }\rangle^{II}_{(c)}=
 {i \over 2}e_1  (i g)^2~\Theta~
{\theta\left(-p_1^+\right)\over p^+_1}
\frac{\theta\left(k^{+}_1-p^+_1\right) }{(k^{+}_1-p^+_1 )
\Delta^{(1a)-}_f}\nonu \times~
~\Lambda_+(k^\prime_{1on})
\gamma^\mu_1~\gamma^+_1 ~\Gamma_1^{\alpha}\Lambda_+(k_{1on})
 \Lambda_+((K_f-k^{\prime}_1)_{on})\Gamma_{2\alpha}\Lambda_+(K_i-k_1)_{on})\ .
 \label{b15jc}
\ee

\subsection{Evaluation of the term $|\overline G_0VG_0 \mathcal{J}^\mu_0(Q)\overline G_0|$}

Let us now discuss the first contribution in  Eq.~(\ref{cvca2}),
$g_0^{-1}|\overline G_0VG_0 \mathcal{J}^\mu_0(Q)\overline G_0|g_0^{-1}$,
depicted by
the diagrams (g) to (i) in Fig.~\ref{cladder2}.  It is worth noting
that the pair contribution is not present due to the conservation of
the plus momentum component. For the sake of simplicity, the two
reducible diagrams, which are present in this first term and
are canceled by the term
$w^{(1)}g_0{{j}^{c(0)\mu}}$ of Eq.~(\ref{cvca2}) in an analogous
way as we have already discussed for diagrams (e) and (d) of Fig.
\ref{cladder1}, are not
shown in Fig.~(\ref{cladder2}).  As we did for the third term in
Eq. (\ref{cvca2}),
we  evaluate first the matrix elements of the relevant part of the
contribution under consideration, i.e.
\be
\Theta~\langle k_1^{\prime +}\vec{k}_{1\perp }^{\prime }|~
\left|
\overline G_0 VG_0\mathcal{J}^\mu_0(1)  \overline G_0\right|~
|k_1^{+}\vec{k}_{1\perp
}\rangle=i e_1 \left(\frac{i g
}{ 2\pi}\right)^2~\Theta~\int dk_1^{\prime -}dk_1^{-}  \nonu
 \frac 1{k_1^{\prime +}(K_f^{+}-k_1^{\prime +})}\frac
{\psla{k}^\prime_{1on}+m_1}{\left(
k_1^{\prime -}- k^{\prime-}_{1on} +i{\varepsilon \over k_1^{\prime +}}
\right) }\frac{(\psla{K}_f-\psla{k}^\prime_1)_{on}+m_2}
{\left( K_f^{-}-k_1^{\prime -}-(K_f-k_1^{\prime})^-_{on}
+i{\varepsilon \over K_f^{+}-k_1^{\prime
+}}\right) } \nonu
\times ~
\frac { \Gamma_1^\alpha\Gamma_{2\alpha}}{(k_1^{\prime+}-p_1^{\prime+} )\left(
k_1^{\prime-}-p_1^{\prime-}-(k_1^\prime-p_1^\prime)^-_{on}+i{\varepsilon \over
k_1^{\prime+}-p_1^{\prime+}}\right) }
\frac {\psla{p}^\prime_{1}+m_1}{p_1^{\prime+}\left(
 p_1^{\prime-}- p^{\prime-}_{1on}  +i{\varepsilon \over p_1^{\prime+}}\right)}~\gamma^\mu_1\nonu
\times ~\frac { 1}{k_1^{+}(K_i^{+}-k_1^{+})}
\frac {\psla{k}_{1on}+m_1}{\left( k_1^{-}-k^{-}_{1on} +i
{\varepsilon \over k_1^{+}}\right) }
\frac {(\psla{K}_i-\psla{k}_1)_{on}+m_2}{
\left( K_i^{-}-k_1^{-}- (K_i-k_1)^-_{on} +i{\varepsilon \over K_i^{+}-k_1^{+}}
\right) }\ , \label{b2n}
\ee
where the intermediate momentum is changed now to $p^{\prime \mu}_1=k^\mu_1+Q^\mu$. In
order to analytically evaluate the integral, in addition to
the  poles $k^-_{1A}$, $k^-_{1B}$, $k^{\prime -}_{1A}$,
$k^{\prime -}_{1B}$, given in Eq. (\ref{pole3}), we have to consider  the following ones
\be
k_{1D}^{ -}+Q^-= p^{\prime -}_{1on}  -i{\varepsilon \over p_1^{\prime +}}
\nonu
k_{1D}^{\prime -}=p_1^{\prime -}+(k_1^\prime-p_1^\prime)^-_{on}-i
{\varepsilon \over
k_1^{\prime+}-p_1^{\prime+}}\ee
with the on-minus-shell definitions given by
\be
p^{\prime -}_{1on}={\vec{p}^{\prime 2}_{1
\perp}+m_1^2 \over p_1^{\prime+}}
\nonu
(k_1^\prime-p_1^\prime)^-_{on}={(\vec{k}^\prime_{1 \perp}-\vec{k}^\prime_{1
\perp})^2+\mu^2 \over k_1^{\prime+}-p_1^{\prime+}}
\ee
As already said in Sect. VII,  we
have always $p_1^{\prime+}>0$ and therefore no pair contribution can be generated
from this term.
In what follows we consider two regions: i)
$k_1^{\prime+}>p_1^{\prime+}=k_1^{+}+Q^+$ and ii) $p_1^{\prime+}=k_1^{+}+Q^+
>k_1^{\prime+}$. Moreover the definition of $p^{\prime \mu}_1$ suggests to start
with the integration over $k^{\prime -}_1$. In this way we have to discuss only three
poles.
\subsubsection{ Diagrams (g) and  (h)  for $k_1^{\prime+}>p_1^{\prime+}
\geq 0$}

From the residue at the pole $k^{\prime -}_{1B}$ one gets: \be
~\Theta~\langle k_1^{\prime +}\vec{k}_{1\perp }^{\prime }|~ \left|
\overline G_0 VG_0\mathcal{J}^\mu_0(1)  \overline G_0\right|~
|k_1^{+}\vec{k}_{1\perp }\rangle_{(g)+(h)}= e_1 \frac{\left(i
g\right)^2 }{ 2\pi}~\Theta~\theta(k_1^{\prime+}-p_1^{\prime+})\nonu
\times~\int dk_1^{-}
 \frac 1{k_1^{\prime +}(K_f^{+}-k_1^{\prime +})}
{ \psla{k}^\prime_{1on}+m_1 \over \left[
K_f^{-}-(K_f-k_1^{\prime})^-_{on}
 - k^{\prime-}_{1on} +i{\varepsilon \over k_1^{\prime +}}
 +i{\varepsilon \over K_f^{+}-k_1^{\prime
+}} \right]}
 ~\left[(\psla{K}_f-\psla{k}^\prime_1)_{on}+m_2\right]
  \nonu
\times ~
\frac { \Gamma_1^\alpha\Gamma_{2\alpha}}{(k_1^{\prime+}-p_1^{\prime+} )\left[
K_f^{-}-(K_f-k_1^{\prime})^-_{on}-p_1^{\prime-}-(k_1^\prime-p_1^\prime)^-_{on}+i{\varepsilon \over
k_1^{\prime+}-p_1^{\prime+}} +i{\varepsilon \over K_f^{+}-k_1^{\prime
+}}\right] }\nonu
\times ~\left [
{\psla{p}^\prime_{1on}+m_1\over p_1^{\prime+}\left(
 p_1^{\prime-}- p^{\prime-}_{1on}  +i{\varepsilon \over p_1^{\prime+}}\right)}
 +{\gamma^+\over 2 p^{\prime+}_1}
\right
]~
 \gamma^\mu_1\nonu
\times ~\frac { 1}{k_1^{+}(K_i^{+}-k_1^{+})}
\frac {\psla{k}_{1on}+m_1}{\left( k_1^{-}-k^{-}_{1on} +i
{\varepsilon \over k_1^{+}}\right) }
\frac {(\psla{K}_i-\psla{k}_1)_{on}+m_2}{
\left( K_i^{-}-k_1^{-}- (K_i-k_1)^-_{on} +i{\varepsilon \over K_i^{+}-k_1^{+}}
\right) }\ , \label{b3n}
\ee
Adopting the same strategy as in Eq. (\ref{fracpar}), one obtains
i) the contribution
illustrated by diagram (g) in Fig. (\ref{cladder2}) and a reducible term
(not shown in  Fig. (\ref{cladder2}) ), ii)
the instantaneous term (h). In order to separate the irreducible contribution
from the reducible one, we exploit the following identity
\be
 \frac 1 {\left[
K_f^{-}-(K_f-k_1^{\prime})^-_{on}-p_1^{\prime-}-(k_1^\prime-p_1^\prime)^-_{on}+i{\varepsilon \over
k_1^{\prime+}-p_1^{\prime+}} +i{\varepsilon \over K_f^{+}-k_1^{\prime
+}}\right]
 }
 \frac 1{\left(p^{\prime -}_1
 - p^{\prime -}_{1on}+i{\varepsilon \over p_1^{\prime+}} \right)}
 =   \nonu
  \left[\frac 1 {\left[
K_f^{-}-(K_f-k_1^{\prime})^-_{on}-p_1^{\prime-}-(k_1^\prime-p_1^\prime)^-_{on}+i{\varepsilon \over
k_1^{\prime+}-p_1^{\prime+}} +i{\varepsilon \over K_f^{+}-k_1^{\prime
+}}\right] }
+ \frac 1{\left(p^{\prime -}_1
 - p^{\prime -}_{1on}+i{\varepsilon \over p_1^{\prime+}
 } \right)}
\right] \nonu \times
\frac1{\left[
K_f^{-}-(K_f-k_1^{\prime})^-_{on}-p^{\prime-}_{1on}-(k_1^\prime-p_1^\prime)^-_{on}+i{\varepsilon \over
k_1^{\prime+}-p_1^{\prime+}} +i{\varepsilon \over K_f^{+}-k_1^{\prime
+}}+i{\varepsilon \over p_1^{\prime+}
 }   \right]}\ ,
\label{fracparn}
\ee
In particular, the term generated by the first contribution in the square
brackets can  integrated over $k^-_1$ taking
the residue at the pole $k^-_{1A}$. Then  one gets
\be
 \langle k_1^{\prime +}\vec{k}_{1\perp }^{\prime }| j^{c(1)\mu}
|k_1^{+}\vec{k}_{1\perp }\rangle_{(g)}=
  i e_1 (2m_1) (i g)^2~\Theta~
\frac{\theta\left(k^{\prime+}_1-p^{\prime+}_1\right)
}{(k^{\prime+}_1-p^{\prime+}_1 )p^{\prime+}_1
\Delta^{(1g)-}_f\Delta^{(1g)-}_i}
 \nonu
\times
\Lambda_+(k^\prime_{1on})\Gamma_1^\alpha\Lambda_+(p^\prime_{1on})\gamma^\mu_1
\Lambda_+(k_{1on})\Lambda_+((K_f-k^\prime_1)_{on})\Gamma_{2\alpha}\Lambda_+((K_i-k_1)_{on})
 \ , \label{b8}
\ee
where the combinations  $\Delta^{(1g)}_f$ and $\Delta^{(1g)}_i$ of
four momenta are
\begin{eqnarray}
&& \Delta^{(1g)}_f=K_f-p^\prime_{1on} - (K_f-k^\prime_1)_{on} -(
k^\prime_1-p^\prime_1)_{on}+i\varepsilon \nonumber \\ &&
\Delta^{(1g)}_i=K_i- k_{1on}-(K_f-k^\prime_1)_{on} -(
k^\prime_1-p^\prime_1 )_{on} +i\varepsilon = \Delta^{(1f)}_i\ . \label{b9}
\end{eqnarray}
{ since $k^\prime_1-p^\prime_1=p_1-k_1$.} The second term in the
square brackets yields a reducible contribution, which is canceled
out by a term from $w^{(1)}g_0{{j}^{c(0)\mu}}$.

The result for the instantaneous diagram (h) can be obtained by
taking the residue at the pole $k^-_{1A}$ in Eq. (\ref{b3n}), i.e.
\begin{eqnarray}
 &&\langle k_1^{\prime +}\vec{k}_{1\perp }^{\prime }| j^{c(1)\mu}
|k_1^{+}\vec{k}_{1\perp }\rangle_{(h)}=
  \frac{i}{2} e_1  (i g)^2~\Theta~
\frac{\theta\left(k^{\prime+}_1-p^{\prime+}_1\right)
}{(k^{\prime+}_1-p^{\prime+}_1 )p^{\prime+}_1\Delta^{(1g)-}_i}
 \nonu
\times
\Lambda_+(k^\prime_{1on})\Gamma_1^\alpha\gamma^+_1\gamma^\mu_1
\Lambda_+(k_{1on})\Lambda_+((K_f-k^\prime_1)_{on})\Gamma_{2\alpha}
\Lambda_+((K_i-k_1)_{on})
 \ , \label{b8inst}
\end{eqnarray}

\subsubsection{ Diagram (i) for $p_1^{\prime+}>k_1^{\prime+}> 0$}
In this kinematical region only an instantaneous contribution is an
irreducible term, the one illustrated by  diagram (i) in Fig.
(\ref{cladder2}). By taking first the residue at the pole $k^{\prime
-}_{1A}$ and then the residue at the pole $k^{ -}_{1B}$, one has \be
 \langle k_1^{\prime +}\vec{k}_{1\perp }^{\prime }| j^{c(1)\mu}
|k_1^{+}\vec{k}_{1\perp }\rangle_{(i)}=  \frac{i}{2} e_1  (i g)^2
~\Theta~\frac{\theta\left(p^{\prime+}_1-k^{\prime+}_1\right)
}{(p^{\prime+}_1-k^{\prime+}_1)p^{\prime+}_1\Delta^{(1i)-}_f}
 \nonu
\times
\Lambda_+(k^\prime_{1on})\Gamma_{1}^{\alpha}\gamma^+_1\gamma^\mu_1
\Lambda_+(k_{1on})\Lambda_+((K_f-k^\prime_1)_{on})\Gamma_{2\alpha}
\Lambda_+((K_i-k_1)_{on})
 \ , \label{b9inst}
\ee
where the three-body denominator is the minus component of
\begin{eqnarray}
\Delta^{(1i)}_f=K_f-k^\prime_{1on}-(K_i-k_1)_{on}-
(p^\prime_1-k^\prime_1)_{on}=\Delta_f^{(1a)} \ ,
\label{delta1i}
\end{eqnarray}

\section{${\mathcal O}(1)$ Current conservation in the ladder Yukawa model }

The operatorial form of the current conservation at  the first order
of the QP expansion, as given by the  expression in
Eq.~(\ref{wtijcn1}), will be explicitly checked in what follows. {
The symbol $\Theta=\theta(k^+_1)
\theta(k^{\prime+}_1)\theta(K^+_i-k^+_1)
\theta(K^+_f-k^{\prime+}_1)$ simplifies the notation. Furthermore,
it is worth noting that the $\Delta$'s, given in  Eqs.~(\ref{b6}),
(\ref{delta1f}), (\ref{b16}), (\ref{b9}) and (\ref{delta1i}), allow
to express the momentum transfer, $Q$, in a convenient way,
depending upon the term we are considering, and   they have only
minus components non vanishing}.

Let us separate two kinematical regions : i) $k^{\prime +}_1>Q^+$ and ii)
$Q^+>k^{\prime +}_1$.
\subsection {$p^{ +}_1=k^{\prime +}_1-Q^+>0$}
In this region one has contribution from diagrams (a), (c$^I$), (f),
(g), (h) in Figs. (\ref{cladder1}) and (\ref{cladder2}) (since $k^{\prime +} -p^{\prime +}_1=p^+_1-k^{ +}_1>0$).   For
diagrams
(a) and (g), the
momentum transfer can be expressed in two different ways, viz
\begin{eqnarray}
 Q=\Delta^{(1a)}_f-\Delta^{(1a)}_i+k^\prime_{1on}-p_{1on} \ ,
 \label{qa}
\\
Q=\Delta^{(1g)}_f-\Delta^{(1g)}_i-k_{1on}+p^\prime_{1on} \ .
\label{g}
 \end{eqnarray}
Therefore, since $\Lambda_+(k^\prime_{1on})(\psla k^\prime_{1on}-\psla
p_{1on})\Lambda_+(p_{1on})= 0$, one has
\begin{equation}
\Lambda_+(k^\prime_{1on})~\Psla{Q}~\Lambda_+(p_{1on})=
\Lambda_+(k^\prime_{1on})\frac{\gamma^+_1}{2}
\Lambda_+(p_{1on})(\Delta^{(1a)-}_f-\Delta^{(1a)-}_i)  \ .
\label{b11}
\end{equation}
Analogously one finds that,
 \begin{equation}
\Lambda_+(p^\prime_{1on})~\Psla{Q}~\Lambda_+(k_{1on})
 =\Lambda_+(p^\prime_{1on})\frac{\gamma^+_1}{2}\Lambda_+(k_{1on})
(\Delta^{(1g)-}_f-\Delta^{(1g)-}_i) \ \ . \label{b12}
\end{equation}
Eqs. (\ref{b11}) and (\ref{b12}) allow to calculate the
four-divergence of the current.

The divergence for the term (a) plus (g) (see
Eqs.~(\ref{b5}) and (\ref{b8})) is given by:
\be
Q\cdot\langle k_1^{\prime +}\vec{k}_{1\perp }^{\prime }|j^{c(1)}
|k_1^{+}\vec{k}_{1\perp }\rangle_{(a)+(g)}=  \Theta ~i e_1 (i
g)^2\theta\left(k^{\prime+}_1-Q^+\right)
   \nonu\times
\left[ \frac{\theta\left(k^{+}_1-p^+_1\right)}{(k^+_1-p^+_1)}
\left(\frac{1}{\Delta^{(1a)-}_i}-\frac{1}{\Delta^{(1a)-}_f}\right)
\Lambda_+(k^\prime_{1on})\frac{m_1
}{p^+_1}\gamma^+_1\Lambda_+(p_{1on})\Gamma_1^{\alpha}\Lambda_+(k_{1on})
   \right. \nonu\left.
 +\frac{\theta\left(k^{\prime+}_1-p^{\prime+}_1\right)
}{(k^{\prime+}_1-p^{\prime+}_1
)}\left(\frac{1}{\Delta^{(1g)-}_i}-\frac{1}{\Delta^{(1g)-}_f}\right)
\Lambda_+(k^\prime_{1on}) \Gamma_1^{\alpha}\Lambda_+(p^\prime_{1on})
\frac{m_1 }{p^{\prime+}_1}\gamma^+_1 \Lambda_+(k_{1on})
 \right] \nonu
 \times  \Lambda_+(k^\prime_{2on})\Gamma_{2\alpha}\Lambda_+(k_{2on})
 \ , \label{b13}
\ee
where we have used   $k_{2on}=(K_i-k_1)_{on}$ and
$k^\prime_{2on}=(K_f-k^\prime_1)_{on}$ to simplify the notation.

Let us consider the divergence of currents (c$^I$) and (i$^I$),
corresponding to the contribution of instantaneous terms in the
kinematical region $k^{\prime+}_1>Q^+>0$. The following relation, obtained by
exploiting i)
the property
$(\gamma^+)^2=0$ and ii) the anticommutation rule $\{ \psla k, \gamma^+\}= 2
k^+$, is useful for
the next formal steps, i.e.
\be
\Lambda_+(k^\prime_{1on})\Psla Q\gamma^+=\Lambda_+(k^\prime_{1on})\left [
\psla k^\prime_{1}-\psla
p_{1}\right ]\gamma^+=\Lambda_+(k^\prime_{1on})\left [
\psla k^\prime_{1on}-\psla
p_{1on}\right ]\gamma^+=\nonu=
\Lambda_+(k^\prime_{1on})\left [m_1-\psla
p_{1on}\right ]\gamma^+=\Lambda_+(k^\prime_{1on})\left [-2p^+_1 + 2m_1
\gamma^+
\Lambda_+(p_{1on})\right ]=\nonu =2p^+_1~\Lambda_+(k^\prime_{1on})\left [-1+
{m_1\over p^+_1 } \gamma^+
\Lambda_+(p_{1on})\right ]
\label{gplus}
\ee
Then, from  Eq.~(\ref{b5inst}), one has
\be Q\cdot\langle k_1^{\prime
+}\vec{k}_{1\perp }^{\prime }|j^{c(1)} |k_1^{+}\vec{k}_{1\perp
}\rangle^I_{(c)}=  \Theta~
ie_1(i
g)^2\frac{\theta\left(k^{\prime+}_1-Q^+\right)\theta\left(k^{+}_1-p^+_1\right)}
{(k^+_1-p^+_1)\Delta_f^{1a-}}
   \nonu \times \Lambda_+(k^\prime_{1on})
\left[ -1 +\frac{m_1}{p^+_1}\gamma^+_1\Lambda_+(p_{1on})
\right]\Gamma^\alpha_1\Lambda_+(k_{1on})\Lambda_+(k^\prime_{2on})
\Gamma_{2\alpha}\Lambda_+(k_{2on}) \ . \label{b13c1}
\ee
For diagram (i), Eq.~(\ref{b9inst}), in this region,
a relation analogous to Eq. (\ref{gplus}) holds, i.e.
\be
\gamma^+~\Psla Q~\Lambda_+(k_{1on})=\gamma^+\left [
\psla p^\prime_{1}-\psla
k_{1}\right ]\Lambda_+(k_{1on})=\gamma^+\left [
\psla p^\prime_{1on}-\psla
k_{1on}\right ]\Lambda_+(k_{1on})=\nonu =
2 p^{\prime +}_1 \left [1-{m_1\over p^{\prime +}_1 }
\Lambda_+(p^\prime_{1on}) \gamma^+\right]
\Lambda_+(k_{1on})
\label{b25}
\ee
Therefore, from Eq. (\ref{b9inst}), one  obtains
\be Q\cdot\langle k_1^{\prime
+}\vec{k}_{1\perp }^{\prime }|j^{c(1)} |k_1^{+}\vec{k}_{1\perp
}\rangle^I_{(i)}=  i~\Theta~e_1(i
g)^2\frac{\theta\left(k^{\prime+}_1-Q^+\right)\theta\left(k^{+}_1-p^+_1\right)}
{(k^+_1-p^+_1)\Delta_f^{(1a)-}}
   \nonu  \times \Lambda_+(k^\prime_{1on})\Gamma^\alpha_1
\left[1
-\Lambda_+(p^\prime_{1on})\frac{m_1}{p^{\prime+}_1}\gamma^+_1
\right]\Lambda_+(k_{1on})\Lambda_+(k^\prime_{2on})
\Gamma_{2\alpha}\Lambda_+(k_{2on}) \ , \label{b13i1}
\ee
 Summing Eqs. (\ref{b13c1}) and
(\ref{b13i1}), we find that:
\be Q\cdot\langle k_1^{\prime
+}\vec{k}_{1\perp }^{\prime }|j^{c(1)} |k_1^{+}\vec{k}_{1\perp
}\rangle^I_{(c)+(i)}=  \imath ~\Theta~ e_1(i
g)^2\frac{\theta\left(k^{\prime+}_1-Q^+\right)\theta\left(k^{+}_1-p^+_1\right)}
{(k^+_1-p^+_1)\Delta_f^{(1a)-}}
   \nonu  \times \Lambda_+(k^\prime_{1on})
\left[\frac{m_1}{p^+_1}\gamma^+_1\Lambda_+(p_{1on})\Gamma^\alpha_1
-\Gamma^\alpha_1\Lambda_+(p^\prime_{1on})
\frac{m_1}{p^{\prime+}_1}\gamma^+_1
\right]\Lambda_+(k_{1on})\Lambda_+(k^\prime_{2on})
\Gamma_{2\alpha}\Lambda_+(k_{2on}) \ . \label{b13ci1}
\ee

The instantaneous terms (f) plus (h), see Eqs. (\ref{b5f}) and (\ref{b8inst})
can be calculated following
the same steps as the ones that lead to Eq.~(\ref{b13ci1}).
Noting that $\Delta_i^{(1g)}=\Delta_i^{(1f)}$, one has:
\be Q\cdot\langle k_1^{\prime
+}\vec{k}_{1\perp }^{\prime }|j^{c(1)} |k_1^{+}\vec{k}_{1\perp
}\rangle_{(f)+(h)}=  \imath ~\Theta~e_1(i
g)^2\frac{\theta\left(k^{\prime+}_1-Q^+\right) \theta\left(
p^+_1-k^{+}_1\right)} {( p^+_1-k^+_1)\Delta_i^{(1g)-}}
   \nonu  \times \Lambda_+(k^\prime_{1on})
\left[\frac{m_1}{p^+_1}\gamma^+_1\Lambda_+(p_{1on})\Gamma^\alpha_1
-\Gamma^\alpha_1\Lambda_+(p^\prime_{1on})\frac{m_1}{p^{\prime+}_1}\gamma^+
\right]\Lambda_+(k_{1on})\Lambda_+(k^\prime_{2on})
\Gamma_{2\alpha}\Lambda_+(k_{2on}) \ . \label{b13fh1}
\ee

Adding Eqs.~(\ref{b13}), (\ref{b13ci1}) and (\ref{b13fh1}) one
gets in the kinematical region $k^{\prime+}_1>Q^+>0$ the result
\be
Q\cdot\langle k_1^{\prime +}\vec{k}_{1\perp }^{\prime
}|j^{c(1)} |k_1^{+}\vec{k}_{1\perp
}\rangle~\theta\left(k^{\prime+}_1-Q^+\right)= \nonu
 \theta\left(k^{\prime+}_1-Q^+\right)~\Theta~
\left[ e_1\Lambda_+(k^\prime_{1on})\frac{m_1
}{p^+_1}\gamma^+_1\Lambda_+(p_{1on})\langle p_1^+\vec{p}_{1\perp
}|w^{(1)}(K_i) |k_1^{+}\vec{k}_{1\perp }\rangle \right. \nonu
\left. - \langle k^{\prime+}_1\vec{k}^\prime_{1\perp
}|w^{(1)}(K_f) |p_1^{\prime+}{\vec p}^\prime_{1\perp }\rangle
e_1\Lambda_+(p^\prime_{1on})\frac{m_1
}{p^{\prime+}_1}\gamma^+_1\Lambda_+(k_{1on})\right]
 \ , \label{b13total}
\ee where the matrix elements of the effective interaction in first
order of QP truncation, for total momentum $K_i$, are given by: {\be
\langle p_1^{ +}\vec{p}_{1\perp }|w^{(1)}(K_i)
|k_1^{+}\vec{k}_{1\perp }\rangle=
 i(i g)^2~\theta(p^+_1)~\theta(k_1^{+})\theta(K^+_i-p^+_1)~\theta(K^+_i-k_1^{+})
 \nonu \times~\left[\frac{\theta\left(
p^+_1-k^{+}_1\right)}{(p^+_1-k^{+}_1)}
\frac{1}{\Delta^{(1g)-}_i}+
\frac{\theta\left(k^{+}_1-p^+_1\right)}{(k^{+}_1-p^+_1)}
\frac{1 }{\Delta^{(1a)-}_i} \right] ~
\Lambda_+(p_{1on})\Gamma^{\alpha}_1\Lambda_+(k_{1on})
\Lambda_+(k^\prime_{2on})\Gamma_{2\alpha}\Lambda_+(k_{2on}) \nonu
\label{w1yukawab}
\end{eqnarray}
For total momentum $K_f$, the matrix elements of $w^{(1)}(K_f)$ can
be easily obtained from the previous expression properly changing
the individual momenta, i.e.
\be
\langle k_1^{\prime+}\vec{k}^\prime_{1\perp }|w^{(1)}(K_f)
|p_1^{\prime +}\vec{p}^\prime_{1\perp }\rangle=
 i(i g)^2~\theta(p^{\prime+}_1)~\theta(k_1^{^\prime+})~
 \theta(K^+_f-p^{\prime+}_1)~\theta(K^+_f-k_1^{^\prime+})\nonu \times
 \left[\frac{\theta\left(
k_1^{\prime+}-p_1^{\prime +}\right)}{(k_1^{\prime+}-p_1^{\prime +})}
\frac{1}{\Delta^{(1g)-}_f}+
\frac{\theta\left(p_1^{\prime +}-k_1^{\prime+}\right)}{(p_1^{\prime +}-k_1^{\prime+})}
\frac{1 }{\Delta^{(1a)-}_f} \right] ~
\Lambda_+(k^\prime_{1on})\Gamma^{\alpha}_1\Lambda_+(p^\prime_{1on})
\Lambda_+(k^\prime_{2on})\Gamma_{2\alpha}\Lambda_+(k_{2on}) \nonu
\label{w1yukawaf}
\ee
Let us remind that $k_1^{\prime+}-p_1^{\prime +}=p^+_1-k^{+}_1$ and $
\theta(p_1^{\prime
+})$ is redundant.} For $\Gamma_\alpha\equiv 1$, i.e. for a scalar
boson exchange, one obtains the expression derived in Ref.
\cite{sales01}.

\subsection {$Q^+ - k^{\prime +}_1>0$}
{ In this  kinematical region diagrams (b) and  (c$^{II}$)
contribute. Furthermore we consider the contribution of diagram (i)
in this region (cf the discussion after Eq. (\ref{b9inst})).} The
divergence of the pair current (see Eq.~(\ref{b15})) is given by:
\begin{eqnarray}
&& Q \cdot \langle k_1^{\prime +}\vec{k}_{1\perp }^{\prime }|
j^{c(1)} |k_1^{+}\vec{k}_{1\perp }\rangle_{(b)}= - i ~\Theta~e_1  (i g)^2
\frac{\theta\left(p^{\prime+}_1-k^{\prime+}_1\right) }{(p^{\prime+}_1-k^{\prime+}_1
)  } \frac{\theta\left(Q^+-k^{\prime+}_1 \right)}{\Delta^{(1a)-}_f}
\nonumber \\ && \times \Lambda_+(k^\prime_{1on}) \frac{
m_1}{p^+_1}\gamma^+_1\Lambda_+(p_{1on})\Gamma_1^{\alpha}\Lambda_+(k_{1on})
 \Lambda_+(k^\prime_{2on})\Gamma_{2\alpha}\Lambda_+(k_{2on})\ ,
 \label{b16b}
\end{eqnarray}
where $Q=\Delta^{(1b)}_\gamma+k^\prime_{1on}-p_{1on}$ has been used (see
Eq.~(\ref{b16}) and note that only $\Delta^{(1b)-}_\gamma$ does not
vanishes).

In the kinematical region under consideration, the divergence of the
current (c) (see Eq.~(\ref{b15jc})) is obtained by using
Eq.~(\ref{gplus}) :
\begin{eqnarray}
&& Q \cdot\langle k_1^{\prime +}\vec{k}_{1\perp }^{\prime }|
j^{c(1)} |k_1^{+}\vec{k}_{1\perp }\rangle^{II}_{(c)}=
 i~\Theta~ e_1  (i g)^2~\Theta~\frac{\theta\left(p^{\prime+}_1-k^{\prime+}_1\right) }
 {(p^{\prime+}_1-k^{\prime+}_1 )  }
 \frac{\theta\left(Q^+-k^{\prime+}_1 \right)}{\Delta^{(1a)-}_f}
\nonumber \\ && \times \Lambda_+(k^\prime_{1on}) \left[ \frac{
m_1}{p^+_1}\gamma^+_1\Lambda_+({p}_{1on})-1\right]
\Gamma_1^{\alpha}\Lambda_+(k_{1on})
 \Lambda_+(k^\prime_{2on})\Gamma_{2\alpha}\Lambda_+(k_{2on})\ .
 \label{divjc}
\end{eqnarray}
The divergence of the current (i), Eq.~(\ref{b9inst}), in this region,
using Eq. (\ref{b25}) results
\begin{eqnarray}
&& Q \cdot \langle k_1^{\prime +}\vec{k}_{1\perp }^{\prime }|
j^{c(1)} |k_1^{+}\vec{k}_{1\perp }\rangle_{(i)}^{II} =
 i e_1  (i g)^2~\Theta~\frac{\theta\left(p^{\prime+}_1-k^{\prime+}_1\right) }
 {(p^{\prime+}_1-k^{\prime+}_1
)  }
 \frac{\theta\left(Q^+-k^{\prime+}_1 \right)}{\Delta^{(1a)-}_f}
\nonumber \\ && \times \Lambda_+(k^\prime_{1on})\Gamma_1^{\alpha}
\left[1- \Lambda_+(p^\prime_{1on})
\frac{m_1}{p^{\prime+}_1}\gamma^+_1 \right]\Lambda_+(k_{1on})
\Lambda_+(k^\prime_{2on})
 \Gamma_{2\alpha}\Lambda_+(k_{2on})
 \ .
 \label{divji}
\end{eqnarray}
where we have used that
$Q=\Delta^{(1g)}_f-\Delta^{(1f)}+p^\prime_{1on}-k_{1on}$.

Finally, adding Eqs. (\ref{b16b}), (\ref{divjc}) and (\ref{divji}),
only the second term in the square bracket of (\ref{divji})
survives, resulting in:
\begin{eqnarray}
&& Q \cdot \langle k_1^{\prime +}\vec{k}_{1\perp }^{\prime }|
j^{c(1)} |k_1^{+}\vec{k}_{1\perp
}\rangle_{(b)+(c)+(i)}~\theta\left(Q^+-k^{\prime+}_1 \right)= - i
e_1 (i g)^2 ~\Theta~
\frac{\theta\left(p^{\prime+}_1-k^{\prime+}_1\right) }{(p^{\prime+}_1-k^{\prime+}_1
)  }
 \frac{\theta\left(Q^+-k^{\prime+}_1
\right)}{\Delta^{(1a)-}_f} \nonumber \\ && \times
\Lambda_+(k^\prime_{1on})\Gamma_1^{\alpha}
 \Lambda_+(p^\prime_{1on}) \Lambda_+(k^\prime_{2on})
 \Gamma_{2\alpha}\Lambda_+(k_{2on})
\Lambda_+(p^\prime_{1on}) \frac{m_1}{p^{\prime+}_1}\gamma^+_1
\Lambda_+(k_{1on}) = \nonu = \langle
k^{\prime+}_1\vec{k}^\prime_{1\perp }|w^{(1)}(K_f)
|p_1^{\prime+}{\vec p}^\prime_{1\perp }\rangle
e_1\Lambda_+(p^\prime_{1on})\frac{m_1
}{p^{\prime+}_1}\gamma^+_1\Lambda_+(k_{1on})
\theta\left(Q^+-k^{\prime+}_1\right)
 \label{b18}\ .
 \label{divjpair}
\end{eqnarray}
This single term corresponds to the interaction calculated at first
order in QP expansion, for total momentum $K_f$, in the considered
kinematical region. The term with $w^{(1)}(K_i)$ vanishes in this
kinematical region, since we have  $Q^+-k^{\prime+}_1=-p^+_1>0$ and therefore
 we are outside the kinematical support of the
interaction, see Eq. (\ref{w1yukawab}).

The total divergence of the current operator is found by adding
Eq.~(\ref{b13total}), (\ref{divjpair}) and $Q\cdot j^{c(o)}$. Then,
by taking into account the definition of the {\it left} and {\it right} LF charge
operator, Eqs. (\ref{lfc9}) and (\ref{lfc10}),  one can  obtain
an expression for
the LF WTI valid for the whole kinematical range
$0<k^{\prime}_1<K^+_f$. This illustrates how one can check the
formal expression for $Q\cdot j^{c(1)}$, given by Eq.~(\ref{wtijcn1}), in the
example  of the Yukawa model in ladder approximation.


\end{document}